\documentclass[journal=jctcce,manuscript=article]{achemso}

\usepackage[version=3]{mhchem} 
\usepackage{threeparttable}
\usepackage{amsmath}
\usepackage{graphicx}
\usepackage{xspace}
\usepackage{latexsym}
\usepackage{subfig}
\usepackage{braket}
\usepackage{cleveref}
\usepackage{threeparttable}
\usepackage{enumerate}
\usepackage{setspace}
\usepackage{pdfpages}
\usepackage{xcolor}

\newcommand*{\bohr}{\ensuremath{a_0}\xspace}
\newcommand*{\mEh}{\ensuremath{mE_h}\xspace}

\title[]{Time-step targeting  time-dependent and dynamical density matrix renormalization group algorithms with ab initio Hamiltonians}

\author{Enrico~Ronca}
\email{enrico.r8729@gmail.com}
\affiliation{Division of Chemistry and Chemical Engineering, California Institute of Technology, Pasadena, CA 91125, USA}
\author{Zhendong Li}
\affiliation{Division of Chemistry and Chemical Engineering, California Institute of Technology, Pasadena, CA 91125, USA}
\author{Carlos A. Jimenez-Hoyos}
\affiliation{Division of Chemistry and Chemical Engineering, California Institute of Technology, Pasadena, CA 91125, USA}
\author{Garnet~Kin-Lic~Chan}
\email{gkc1000@gmail.com}
\affiliation{Division of Chemistry and Chemical Engineering, California Institute of Technology, Pasadena, CA 91125, USA}

\abbreviations{}
\keywords{}

\begin{document}


\begin{abstract}
We study the dynamical density matrix renormalization group (DDMRG) and time-dependent density matrix renormalization group (td-DMRG)
algorithms in the ab initio context, to compute dynamical correlation functions of correlated systems. We analyze the 
strengths and weaknesses of the two methods in small model problems, and propose two simple improved formulations, DDMRG$^{++}$ and td-DMRG$^{++}$,
that give  increased accuracy at the same bond dimension, at a nominal increase in cost. We apply DDMRG$^{++}$ to
obtain the oxygen core-excitation energy in the water molecule in a quadruple-zeta quality basis, which allows us to estimate the
remaining correlation
error in existing coupled cluster results. Further, we use DDMRG$^{++}$ to compute the local density of states and gaps, and td-DMRG$^{++}$ to compute
the complex polarization function, in linear hydrogen chains with up to 50 H atoms, to study metallicity and delocalization as a function of bond-length.
\end{abstract}

\titlepage

\maketitle

\section{Introduction}
\label{sec:Introduction}

The calculation of dynamical quantities is essential for the interaction between theory and experiment.
Most commonly, dynamical quantities such as the single-particle Green's function or optical absorption are considered in the linear response regime.
In the frequency domain, the linear response of a wavefunction to a field
can be written as the second derivative of a Lagrangian~\cite{mcweeny1992methods,helgaker2012recent} and frequency-domain response
theory in quantum chemistry has closely followed the theory of analytic energy derivatives, similar to that in structural optimization.
Thus algorithms exist to compute dynamical correlation functions from Hartree-Fock~\cite{dalgarno1966time}, density functional theory~\cite{casida1995response}, configuration interaction~\cite{koch1991analytical},
coupled cluster~\cite{koch1990coupled}, and Jastrow-Slater wavefunctions~\cite{mussard2017time,zhao2016equation} amongst others, using analytic derivative techniques.
Dynamical quantities can also be calculated in the time-domain. Here, quantum chemical methods typically formulate the equation of motion for the wavefunction 
 from the Dirac-Frenkel (time-dependent) variational principle~\cite{krause2007molecular,li2005time,hoodbhoy1978time}. 
Both kinds of algorithms can be found implemented in many modern quantum chemistry codes.

Dynamical quantities have also been studied with density matrix renormalization group (DMRG) or matrix
product state (MPS) wavefunctions. 
Here a wide range of numerical algorithms have been explored.
In the frequency domain, the first dynamical correlation functions
were computed in a fixed linear space of DMRG renormalized states (i.e. by optimizing a single tensor in the MPS)~\cite{hallberg1995density}.
Subsequent algorithms, such as the dynamical DMRG (DDMRG)~\cite{JeckelmannPhysRevB2002,RamaseshaSynthMet1997,PatiPhysRevB1999,KuhnerPhysRevB1999} or analytic DMRG response theory~\cite{DorandoJChemPhys2009,NakataniJChemPhys2014},  further considered the response of the DMRG renormalized basis (i.e.
all tensors in the MPS).  DDMRG is widely used as a benchmark method for DMRG dynamical correlation functions, but unlike
the analytical DMRG response theory does not correspond to a true derivative of a Lagrangian. The analytic DMRG response theory is equivalent to the 
later ``tangent space'' formulations of DMRG dynamical correlation functions~\cite{haegeman2013post}.

Time-propagation has also been investigated in conjunction with DMRG wavefunctions. 
Although a wide variety of time-propagation algorithms have been discussed~\cite{CazalillaPhysRevLett2002,VidalPhysRevLett2003,DaleyJStatMec2004,
 FeiguinPhysRevB2005,WhitePhysRevLett2004,kinder2011analytic,haegeman2011time,ZaletelPhysRevB2015}, some, such as time-evolving block decimation~\cite{VidalPhysRevLett2003}, are specialized to Hamiltonians
with short-range interactions on a 1D lattice. For quantum chemistry, it is necessary
to work with long-range interactions, and one of the early time-dependent DMRG (td-DMRG) algorithms
that supported such Hamiltonians was the time-step targeting time-dependent DMRG method~\cite{FeiguinPhysRevB2005}.
There have also been many other important developments in time-dependent DMRG 
which we do not discuss here, including translating time-propagation algorithms such as Chebyshev expansion and Krylov space techniques to work with MPS~\cite{HolznerPhysRevB2011,GanahlPhysRevB2014,WolfPhysRevB2015}, 
analytic time-propagation  using the time-dependent variational principle~\cite{kinder2011analytic,haegeman2011time}, and matrix product operator  representations of the time-evolution operator with improved
global time-step error~\cite{ZaletelPhysRevB2015}.

In the current work, we explore frequency-dependent and time-dependent DMRG algorithms for
dynamical quantities  to better understand the behaviour and applicability of these algorithms
in the ab initio DMRG context~\cite{WhitePhysRevLett1992,WhitePhysRevB1993,WhiteJChemPhys1999,DaulIntJQuantumChem2000,
MitrushenkovJChemPhys2001,ChanJChemPhys2002,LegezaPhysRevB2003,ChanJChemPhys2004,SchollwockRevModPhys2005,MoritzJChemPhys2006,MartiJChemPhys2008,
MartiZPhysChem2010,SchollwockAnnPhys2011,SharmaJChemPhys2012,WoutersEurPhysJD2014,OlivaresAmayaJChemPhys2015}. 
There has been relatively little work computing  ab initio dynamical quantities with DMRG. 
Earlier work in our group compared dynamical DMRG and analytic DMRG response theory for computing
frequency dependent polarizabilities~\cite{DorandoJChemPhys2009}. Subsequent investigations exploited the analogy between
the analytic DMRG response theory and the random phase approximation to obtain DMRG excitation energies
and RPA-like correlation energy contributions for some small molecules~\cite{NakataniJChemPhys2014}. 
To our knowledge, time-dependent DMRG techniques have not yet been explored with ab initio Hamiltonians, although some studies have been
carried out with model Hamiltonians of conjugated systems~\cite{ma2009dynamical}.

We will focus here on the dynamical DMRG (DDMRG) and time-step targeting time-dependent DMRG (td-DMRG) methods. We concentrate on these techniques
rather than the analytic DMRG response or other time-dependent formulations for two reasons. First, DDMRG and td-DMRG are
simple to implement in existing DMRG codes (and are thus commonly used in applications outside of quantum chemistry). Second, our
work on analytic DMRG theories showed that the quality of the response functions is tied to the similarity between the excited states and the ground-state,
thus excited states with quite different entanglement structure to the ground-state are poorly described except using large bond dimensions~\cite{NakataniJChemPhys2014}. Since the primary
purpose of DMRG in quantum chemistry is to describe strongly correlated systems where we can often find states of
different electronic character at low energies, it is of interest to work with techniques which treat
states with different character in a relatively balanced way. This is the case with DDMRG and td-DMRG methods, which 
treat the response wavefunction or time-evolved state on an equal footing with the ground-state or initial state.
In particular, we will introduce two small improvements to the techniques, that we call DDMRG$^{++}$ and td-DMRG$^{++}$. Although
the change to the algorithms is small and easy to implement within existing DDMRG and td-DMRG codes, the
subsequent improvement in accuracy and concomitant savings in cost is significant. 

The outline of the paper is as follows.
In the section~\ref{sec:Theory} we give a brief overview of linear response theory
dynamical correlation functions as well as frequency-dependent and time-dependent
algorithms to compute Green's functions. We subsequently give some
background on DMRG and MPS, before discussing the detailed theory of the DDMRG and td-DMRG algorithms,
as well as their DDMRG$^{++}$ and td-DMRG$^{++}$ improvements. 
In section~\ref{sec:results} we benchmark DDMRG$^{++}$ and td-DMRG$^{++}$ on small systems which
can be exactly treated by full configuration interaction. We next use DDMRG$^{++}$ to compute
the O $1s$ core excitation energy of the water molecule in realistic basis sets. Finally, we 
use DDMRG$^{++}$ to compute the LDOS and gaps of hydrogen chains up to \ce{H_{50}} within a minimal basis,
and further use td-DMRG$^{++}$ to obtain the complex polarization function to characterize the metallicity of the
ground-state as a function of bond-length. We finish with some perspectives in section~\ref{sec:conclusions}.

\section{Theoretical Methods}
\label{sec:Theory}
\subsection{Linear response}
\label{sec:lin_resp}

When the applied fields are not too strong, linear response theory underpins spectroscopy. We
briefly recap the essentials here. Consider a system in an initial eigenstate $\Psi_0$ of a Hamiltonian
$\hat{H}_0$, and consider a time-dependent perturbation $f(t) \hat{V}(t)$, where
$f(t)$ is the field strength. The linear response of the observable $\hat{O}$ is given by
\begin{equation}\label{eq:1_expec_var}
\delta\langle\Psi_0|\hat{O}(t)|\Psi_0\rangle =
\int_{-\infty}^{t}dt' \chi(t-t') f(t')
\end{equation}
where $\hat{O}(t) = e^{i\hat{H_0}t} \hat{O} e^{-i\hat{H_0}t}$ and
the Kubo formula\cite{KuboJPhysSocJap1957} for the generalized susceptibility $\chi(t-t')$ is:
\begin{equation}\label{eq:kubo}
\chi(t-t')=-{i}\theta(t-t')\langle\Psi_0|[\hat{O}(t),\hat{V}(t')]|\Psi_0\rangle.
\end{equation}
The frequency dependent susceptibility is:
\begin{eqnarray}
\chi(\omega) &=&
\int_{-\infty}^\infty d(t-t') e^{i\omega (t-t')}\chi(t-t')\nonumber\\
&=&
\sum_m\frac{\langle\Psi_0|\hat{O}|\Psi_m\rangle\langle\Psi_m|\hat{V}|\Psi_0\rangle}{\omega-(E_m-E_0)+i\eta}
-\sum_n\frac{\langle\Psi_0|\hat{V}|\Psi_n\rangle\langle\Psi_n|\hat{O}|\Psi_0\rangle}{\omega-(E_0-E_n)+i\eta},\label{eq:freq_susc}
\end{eqnarray}
where $\eta$ is a infinitesimal positive number, $\Psi_{m(n)}$ are excited states of the system,
$E_{m(n)}$ are the associated eigenvalues.
The imaginary part of the susceptibility is the spectral function, which is proportional to the rate
of absorption of the applied field~\cite{ColemanBook2015},
\begin{equation} \label{eq:spectral}
S(\omega)=-\frac{1}{\pi}\mathrm{Im}\chi(\omega).
\end{equation}

Different spectroscopies are described by different combinations of the operators $\hat{O}$ and $\hat{V}$.
For example, optical spectroscopy is described by $\hat{O}, \hat{V} = \hat{\mu}$, where $\hat{\mu}$ is the dipole operator.
Likewise,  photoelectron spectroscopy can be
described by the retarded Green's function,
\begin{eqnarray}
G^R_{ij}(t-t')=
-i\theta(t-t')\langle\Psi_0^N|[a_i(t),a_j^\dagger(t')]_+|\Psi_0^N\rangle,\label{GFRdef}
\end{eqnarray}
where $\hat{O}, \hat{V} = a_i/a_j^\dagger$ respectively, $a_i^{(\dag)}$
are creation/annihilation operators, and $[\hat{A},\hat{B}]_+=\hat{A}\hat{B}+\hat{B}\hat{A}$ is
the anticommutator. Its Lehmann representation reads
\begin{eqnarray}
G^R_{ij}(\omega)
=
\sum_m\frac{\langle\Psi_0^N|a_i|\Psi_m^{N+1}\rangle\langle\Psi_m^{N+1}|a_j^\dagger|\Psi_0^N\rangle}
{\omega-(E_m^{N+1}-E_0^N)+i\eta}
+
\sum_n\frac{\langle\Psi_0^N|a_j^\dagger|\Psi_n^{N-1}\rangle\langle\Psi_n^{N-1}|a_i|\Psi_0^N\rangle}
{\omega-(E_0^N-E_n^{N-1})+i\eta}.\label{GFR}
\end{eqnarray}
The spectral function or density of states (LDOS) becomes
\begin{eqnarray}
S_{ij}(\omega)&=&-\frac{1}{\pi}\mathrm{Im}G^R_{ij}(\omega)\nonumber\\
&=&
\sum_m
\langle\Psi_0^N|a_i|\Psi_m^{N+1}\rangle\langle\Psi_m^{N+1}|a_j^\dagger|\Psi_0^N\rangle
\delta(\omega-(E_m^{N+1}-E_0^N))\nonumber\\
&+&
\sum_n
\langle\Psi_0^N|a_j^\dagger|\Psi_n^{N-1}\rangle\langle\Psi_n^{N-1}|a_i|\Psi_0^N\rangle
\delta(\omega-(E_0^N-E_n^{N-1})).
\end{eqnarray}
In this work, we will focus on the Green's function and density of states as measured by photoelectron spectroscopy, but the
formalism  can  easily be extended to other spectroscopies.

\subsection{Frequency and time-domain calculations of Green's functions}
\label{sec:general_gf}

We can obtain equivalent information on the linear response in the frequency and in the time-domain.
We now discuss general strategies to compute the Green's function in these two settings.
Notice that the Green's function has two contributions, see Eq. \eqref{GFR}.
The first part corresponds to the electron addition (EA) component of the Green's function, while the second part corresponds to the electron removal (IP) one.
Computationally, we can compute the two pieces separately. Below we present explicit formulae only for the IP part, and
analogous derivations apply to the EA part.

Formally, the frequency ($\omega$)-dependent IP Green's function matrix element $G_{ij}(\omega)$ \eqref{GFR} can be
rewritten as,
\begin{equation}\label{eq:omegaGF}
G_{ij}(\omega)=\langle\Psi_0|a_j^{\dagger}\frac{1}{\omega+\hat{H}_0-E_0+i\eta}a_i|\Psi_0\rangle.
\end{equation}
It is convenient to compute the Green's function from the response equation:
\begin{equation}\label{eq:lin_eq_prob}
[\hat{H}_0-E_0+\omega+i\eta]|c(\omega)\rangle = a_i|\Psi_0\rangle
\end{equation}
where $c(\omega)$ is referred to as the correction vector~\cite{PatiPhysRevB1999,KuhnerPhysRevB1999},
such that the Green's function element is the expectation value
\begin{equation}\label{eq:GF_expec}
G_{ij}(\omega)=\langle\Psi_0|a_j^{\dagger}|c(\omega)\rangle.
\end{equation}

Using real arithmetic, we solve for the real ($|X(\omega)\rangle=\mathrm{Re}|c(\omega)\rangle$) and imaginary parts
($|Y(\omega)\rangle=\mathrm{Im}|c(\omega)\rangle$) of the correction vector separately.
To compute the imaginary part from the equation,
\begin{equation}\label{eq:imag_c_eq0}
[(\hat{H}_0-E_0+\omega)^2+\eta^2]|Y(\omega)\rangle=-\eta a_i|\Psi_0\rangle,
\end{equation}
we can in general minimize the Hylleraas-like functional~\cite{JeckelmannPhysRevB2002},
\begin{equation}\label{eq:imag_c_eq}
\mathcal{L}[Y(\omega)]=
\langle Y(\omega)|[(\hat{H}_0-E_0+\omega)^2+\eta^2]|Y(\omega)\rangle+2\eta\langle Y(\omega)| a_i|\Psi_0\rangle.
\end{equation}
From the imaginary part, the real part can be obtained as:
\begin{equation}\label{eq:real_c_eq}
|X(\omega)\rangle=-\frac{\hat{H}_0-E_0+\omega}{\eta}|Y(\omega)\rangle.
\end{equation}

In the time ($t$) domain the IP part of the Green's function \eqref{GFRdef} is written as:
\begin{equation}\label{eq:final_timeGF}
G_{ij}(t-t') = -i\theta(t-t')\langle\Psi_0|a_j^{\dagger}e^{i(\hat{H}-E_0)(t-t')}a_i|\Psi_0\rangle.
\end{equation}
The steady state Green's function is obtained at sufficiently long time $t \to \infty$.
From this, the frequency dependent Green's function \eqref{eq:omegaGF} can be obtained by Fourier transform,
\begin{equation}\label{eq:ft}
G_{ij}(\omega) = \int_{-\infty}^{\infty} d(t-t') e^{i\omega (t-t')}G_{ij}(t-t').
\end{equation}
Eq.~(\ref{eq:final_timeGF}) can be evaluated by a real-time propagation of an initial state ($a_i|\Psi_0\rangle$).
There are many methods to carry out the time-propagation~\cite{VidalPhysRevLett2003,CazalillaPhysRevLett2002,DaleyJStatMec2004,
FeiguinPhysRevB2005,WhitePhysRevLett2004,ZaletelPhysRevB2015}; in this work we use the simple Runge-Kutta (RK4) algorithm,
which requires calculating four vectors:
\begin{align}\label{eq:RK4_vectors}
&|r_1\rangle = \tau(\hat{H}_0-E_0)|\Psi(t)\rangle\nonumber\\
&|r_2\rangle = \tau(\hat{H}_0-E_0)[|\Psi(t)\rangle+1/2|r_1\rangle]\nonumber\\
&|r_3\rangle = \tau(\hat{H}_0-E_0)[|\Psi(t)\rangle+1/2|r_2\rangle]\nonumber\\
&|r_4\rangle = \tau(\hat{H}_0-E_0)[|\Psi(t)\rangle+|r_3\rangle]
\end{align}
where $|\Psi(t)\rangle$ is the
wavefunction at the initial time-step and $\tau$ is the time-step.
From these four vectors the state at time $t+\tau$ can then be obtained as:
\begin{equation}\label{eq:RK4_prop}
|\Psi(t+\tau)\rangle \approx \frac{1}{6}[|r_1\rangle+2|r_2\rangle+2|r_3\rangle+|r_4\rangle].
\end{equation}
The total accumulated time-step error is $O(\tau^4)$.

We  will next see how to translate these general expressions to compute Green's functions in
the language of DMRG.

\subsection{DMRG and MPS}
To lay some foundations for the time-dependent algorithms, we recall the main ideas of DMRG and Matrix Product States (MPS).
For details, the reader is referred to the recent reviews, see Refs.~\citenum{SchollwockAnnPhys2011,SzalayIntJQuantumChem2015} and \citenum{ChanJChemPhys2016}.
The MPS is the underlying variational wavefunction ansatz used in DMRG algorithms, and is a non-linear
parametrization for the wave function of the form:
\begin{eqnarray}\label{FSMPS}
\ket{\Psi}
= \sum_{\{ n_k \}, \{\alpha_k\}} A^{n_1}_{\alpha_1}[1]A^{n_2}_{\alpha_1\alpha_2}[2]\cdots A^{n_K}_{\alpha_{K-1}}[K]\ket{n_1 n_2 \ldots n_K}
\end{eqnarray}
where $\ket{ n_1 n_2 \cdots n_K}$ is an occupation vector in the Fock space, and $A^{n_k}[k]$ is
an $M \times M$ matrix of numbers, while $A^{n_1}[1]$ and $A^{n_K}[K]$ are $1\times M$ and $M\times 1$ vectors. For a given
occupancy vector, the product of matrices (with vectors for the leftmost and rightmost sites) yields the scalar wavefunction amplitude.
$M$ is the bond dimension (also known as the number of renormalized states) of the DMRG wavefunction. As $M \to \infty$ (or
in a finite Fock space $\mathcal{F}$, $M \to \sqrt{\mathrm{dim}\mathcal{F}}$) the MPS becomes an exact representation
of any state.

In the most general sense, the DMRG algorithm provides a way to determine the tensors in the MPS
one by one from $A^{n_1}[1]$ to $A^{n_K}[K]$ (holding all other tensors fixed at each step)
from the variational principle, or equivalently the minimization of the Lagrangian,
\begin{eqnarray}
\mathcal{L}[\Psi]=\langle\Psi|\hat{H}|\Psi\rangle - E (\langle\Psi|\Psi\rangle-1).\label{LagE}
\end{eqnarray}
One such determination of all the tensors (going forwards and backwards) is called a \emph{sweep}.
Note that the tensors are not unique because of the product form of the MPS;
gauge matrices $G G^{-1}$ may be inserted in between the tensors while keeping the state invariant.
To properly condition the optimization, when optimizing the $k$th tensor, we use the so-called mixed canonical gauge around site $k$:
\begin{align}\label{eq:mixed}
  \Psi^{n_1n_2\cdots n_K}
  &=\sum_{\{\alpha_k\}} L^{n_1}_{\alpha_1}[1]
  \cdots L^{n_{k-1}}_{\alpha_{k-2}\alpha_{k-1}}[k-1] C^{n_k}_{\alpha_{k-1}\alpha_k}[k]
  R^{n_{k+1}}_{\alpha_k \alpha_{k+1}}[k+1] \cdots R^{n_K}_{\alpha_{K-1}}[K]
\end{align}
where the tensors to the left and right of $k$ satisfy the orthogonality conditions
respectively:
\begin{align}\label{eq:orthocond}
  \sum_{n_k} {L^{n_k}}^T L^{n_k} &=1 \nonumber\\
  \sum_{n_k} {R^{n_k}} {R^{n_k}}^T &=1.
\end{align}
Because of the orthogonality conditions, the $L$ and $R$ tensors collectively define orthogonal sets of many-particle renormalized bases, recursively,
\begin{eqnarray}\label{eq:rbasis}
  \ket{l_{\alpha_{k-1}}} &=  &\sum_{  n_1 \cdots n_k  } (L^{n_1}[1] L^{n_2}[2] \cdots L^{n_{k-1}}[k-1])_{\alpha_{k-1}} |n_1 \cdots n_{k-1}\rangle \nonumber\\
  \ket{r_{\alpha_k}} &=  &\sum_{  n_{k+1} \cdots n_K } (R^{n_{k+1}}[k+1] R^{n_{k+2}}[k+2] \cdots R^{n_K}[K])_{\alpha_k} |n_{k+1} \cdots n_K\rangle
\end{eqnarray}
and the MPS wavefunction may be equivalently written in the space of these renormalized states as:
\begin{equation}\label{eq:dmrg_wf}
  |\Psi[k]\rangle = \sum_{\alpha_{k-1} n_k \alpha_k} C^{n_k}_{\alpha_{k-1} \alpha_k}[k] \ket{l_{\alpha_{k-1}} n_k r_{\alpha_k}},
\end{equation}
where the symbol $[k]$  indicates that the wave function is in the mixed canonical form at site $k$.
At each site in a DMRG sweep one performs several operations: constructing the renormalized bases and the renormalized operators in these bases
at each site $k$ (\emph{blocking});
determining the site wavefunction $C^{n_k}_{\alpha_{k-1} \alpha_k}[k]$ (\emph{solving}); and transforming
all quantities to the canonical form of the next site (\emph{decimation}).

For example, in the ground-state DMRG algorithm, at each site $k$, we build the renormalized site Hamiltonian ($\hat{H}[k]$)
by projecting the Hamiltonian ($\hat{H}$) into the renormalized basis of the site:
\begin{equation}\label{eq:site_ham}
  \hat{H}[k] = P[k] \hat{H} P[k]
\end{equation}
where $P[k]=\sum_\alpha \ket{m[k]_\alpha} \bra{m[k]_\alpha}$ projects into the basis  $\{ \ket{m[k]_\alpha} \} = \{ \ket{l_{\alpha_{k-1}} n_k r_{\alpha_k}} \}$.
Then, Eq. \eqref{LagE} becomes a quadratic function in $C^{n_k}_{\alpha_{k-1} \alpha_k}[k]$.
We then solve for the ground-state of $\hat{H}[k]$ through:
\begin{equation} \label{eq:gs_dmrg}
  \hat{H}[k]|\Psi[k]\rangle = E|\Psi[k]\rangle,
\end{equation}
which amounts to a standard eigenvalue problem for
$C^{n_k}_{\alpha_{k-1} \alpha_k}[k]$ in Eq. \eqref{eq:dmrg_wf}.
The final step is to transform
all quantities to the mixed canonical gauge at the neighbouring site.
We do so by building the density matrix $\Gamma[k](C^{n_k}[k])$
in the blocked basis $\{\ket{l_{\alpha_{k-1}} n_k}\}$
with matrix elements:
\begin{equation}\label{eq:densmat}
  \Gamma[k]_{{\alpha_{k-1}} n_k, {\alpha'_{k-1}}n_k'} = (C[k]C[k]^\dag)_{{\alpha_{k-1}} n_k, {\alpha'_{k-1}}n_k'},
\end{equation}
where $C[k]$ is the reshaped matrix $C_{{\alpha_{k-1}}n_k,{\alpha_k}}[k]$ from the tensor $C^{n_k}_{\alpha_{k-1} \alpha_k}[k]$.
The $M$ eigenvectors of the $\Gamma[k]$ with the largest eigenvalues form a matrix with elements $L[k]_{{\alpha_{k-1}} n_k, {\alpha_k}}$; when reshaped to $L[k]^{n_k}_{{\alpha_{k-1}}, {\alpha_k}}$
this becomes the tensor that replaces $C^{n_k}[k]$ in the MPS.
A guess for the site-wavefunction at site $k+1$ can be obtained by
transforming $C^{n_k}[k]$\cite{ChanJChemPhys2002}:
\begin{equation}\label{eq:general_trans}
  C^{n_{k+1}}_{{\alpha_{k}}{\alpha_{k+1}}}[k+1] =
  \sum_{{\alpha'_{k}}}
  (L[k]^\dagger C[k])_{{\alpha_k}{\alpha'_k}}
  R^{n_{k+1}}_{{\alpha'_k}{\alpha_{k+1}}}[k+1],
\end{equation}
where both $L[k]$ and $C[k]$ are the matrix versions of the site tensors, respectively.

In many DMRG algorithms, one is interested in simultaneously representing multiple states $\ket{\Psi_i}$ as matrix product states.
It can be convenient computationally to constrain these MPS such that different states use the same renormalized bases at each site; then
each state is distinguished only by its respective site wavefunction $C^{n_k}[k]_i$. Such algorithms are known as
state-averaged algorithms. To construct the common renormalized bases at each site, one transforms
bases from site to site via the ``state-averaged'' density matrix:
\begin{equation}\label{stateave_densmat}
  \Gamma[k] = \sum_i w_i \Gamma[k]_i(C^{n_k}[k]_i)
\end{equation}
where $w_i$ are weights and $\Gamma[k]_i$ are the density matrices of the individual states entering into the average
computed using Eq.~\eqref{eq:densmat}. In this case, the density matrix has more than $M$ non-zero eigenvalues and
the transformation from site to site does not precisely preserve the states unless $M \to \infty$.
For finite $M$ this requires choosing a site at which to compute observables. In our case,
we report observables calculated at the middle of the sweep, although other choices are possible.

Finally, we mention that in the following sections, the action of an operator $\hat{O}$ on an MPS $\hat{O}|\Psi_0\rangle$ will be frequently encountered (e.g. $a_i|\Psi_0\rangle$ on the right hand side of
Eq. \eqref{eq:imag_c_eq0}). In certain cases, it is necessary to reduce the bond dimension of the state $\hat{O}|\Psi_0\rangle$, for example
in the variational compression used in the benchmark td-DMRG(G) algorithm below, or if one needs to use a smaller bond dimension in the
DDMRG$^{++}$ calculation than in the ground-state DMRG calculation.
The reduction in bond dimension can  in general be achieved via a variational compression by constructing the
``least-squares'' functional,
\begin{eqnarray}
\mathcal{L}[\Psi]=\langle\Psi-\hat{O}\Psi_0|\Psi-\hat{O}\Psi_0\rangle.
\end{eqnarray}
Similar to the minimization of Eq.~\eqref{LagE} for the ground state, the MPS representation 
$|\Psi\rangle$ for $\hat{O}|\Psi_0\rangle$ can be obtained by minimizing this functional using analogous DMRG sweeps. 
The only difference is that instead of solving an eigenvalue problem \eqref{eq:gs_dmrg}, a linear equation needs to
be solved at each site $k$, whose solution in the mixed canonical form is simply given by
the local projection $|\Psi[k]\rangle =P[k]\hat{O}|\Psi_0\rangle$.

\subsection{DDMRG$^{++}$}
\label{subsec:omegaDMRG}

We now discuss how to determine the frequency-dependent Green's function
using MPS and the dynamical DMRG (DDMRG) algorithm. As discussed earlier,
the DDMRG algorithm has proven to be one of the most accurate methods
to compute Green's functions and other frequency dependent correlation functions within a MPS representation. We earlier studied its
performance for chemical problems in Ref.~\citenum{DorandoJChemPhys2009}.
First, we recap the algorithm and then describe a modification to improve its formal properties and accuracy
which we term DDMRG$^{++}$.

The basic path to transcribe the equations
in Sec.~\ref{sec:general_gf} into a DMRG algorithm
is to translate each equation to the wavefunctions and operators at each site of the DMRG sweep.
The states and operators are then expressed in the renormalized basis $\{ \ket{m[k]_\alpha} \}$ at site $k$.
The simplest choice is to work with
a state-averaged formalism, such that all MPS share the same renormalized basis at each site.
In the standard DDMRG algorithm, we first solve equation~\eqref{eq:gs_dmrg} at site $k$ for the
ground-state wavefunction $|\Psi_0[k]\rangle$. Then, we solve for the correction vector $|c[k]\rangle$ at each site, where in Eq.~(\ref{eq:imag_c_eq})
we additionally use the projected quantities $\hat{H}_0[k]$ and $a_k[k]|\Psi_0[k]\rangle$.
Note that the Hylleraas functional of Eq.~(\ref{eq:imag_c_eq}) involves
the square of the Hamiltonian operator, and $P[k] \hat{H}_0^2 P[k] \neq \hat{H}_0[k]^2$, but this approximation becomes
exact in the limit $M \to \infty$.
To ensure that all states continue to share the same renormalized basis throughout the sweep, we construct the
density matrix for the decimation using equally weighted contributions from $|\Psi_0[k]\rangle$, $a_i^{(\dagger)}[k]|\Psi_0[k]\rangle$, $|X(\omega)[k]\rangle$,  $|Y(\omega)[k]\rangle$.

The accuracy of the DDMRG procedure is controlled by the bond dimension $M$. This governs the quality of the representation
of the states such as $|\Psi_0[k]\rangle$ and $|c(\omega)[k]\rangle$, as well as the quality of the resolution of the identity approximation for $\hat{H}_0^2$.
In a finite system, the imaginary factor $i\eta$ can be chosen arbitrarily, but a smaller $\eta$ leads to more iterations
in minimizing the Hylleraas functional, and a larger bond dimension is needed to represent $|c(\omega)[k]\rangle$ accurately.

Despite the established power of the DDMRG,
there are a few drawbacks to the algorithm, some of which we discussed in Ref.~\citenum{DorandoJChemPhys2009}.
These stem from the use of the state-averaged formalism, which means that
some accuracy in the representation of each state is lost for a given bond dimension $M$.
For example, the ground-state wavefunction in DDMRG for a given $M$ is less accurate than that obtained in the standard ground-state DMRG algorithm.
A related side-effect is that even after completing a ground-state DMRG calculation, it is necessary
to re-optimize the (worse) ground-state in DDMRG to accommodate the new renormalized basis.
For these reasons, we have modified the original dynamical DMRG algorithm to avoid these problems; we term
the modified algorithm, DDMRG$^{++}$. Roughly speaking, we allow each of the states appearing in the
response equation to be an independent MPS (and thus to generate
	its own renormalized basis at each site $k$). More precisely, to avoid
complex MPS tensors, we keep $|\Psi_0\rangle$, $a_i^{(\dagger)}|\Psi_0\rangle$ as
independent MPS, and the pair $|X(\omega)\rangle$, $|Y(\omega)\rangle$
 are represented within a common renormalized basis.
This means that we can re-use the solution of a ground-state DMRG sweep as $|\Psi_0\rangle$ and there
 is no loss of accuracy in the ground-state representation during the DDMRG$^{++}$ sweeps.
The modified DDMRG$^{++}$ scheme can be summarized as follows:
\begin{itemize}
\item A ground-state DMRG calculation is carried out to obtain $E_0$ and the MPS $|\Psi_0\rangle$.
\item We compute a separate MPS, $a_i^{(\dagger)}|\Psi_0\rangle$.
\item We carry out the DDMRG$^{++}$ sweep where we minimize the functional in Eq.~(\ref{eq:imag_c_eq})  at each site $k$ using
  the conjugate gradient algorithm. At site $k$, this gives the correction vectors $|X(\omega)[k]\rangle$, $|Y(\omega)[k]\rangle$.
\item  $|X(\omega)[k]\rangle$ and $|Y(\omega)[k]\rangle$ are averaged
  in the density matrix, which is used to transform all quantities to the next site in the sweep,
  and the sweeps are iterated until convergence.
\end{itemize}

\subsection{td-DMRG$^{++}$}
\label{subsec:RTDMRG}

The time-dependent DMRG (td-DMRG) algorithm that we will discuss was introduced by Feiguin and White and belongs to the
family of adaptive time-dependent DMRG (td-DMRG) methods.
It is based on the 4$^{th}$ order Runge Kutta (RK4) algorithm described in Sec.~\ref{sec:general_gf}.
The advantage of this td-DMRG algorithm is that it is quite simple to implement for Hamiltonians with non-local interactions (as
relevant for quantum chemistry) within a standard DMRG program.
We first describe Feiguin and White's td-DMRG algorithm and then describe
an improvement to this algorithm that we will call td-DMRG$^{++}$.

As discussed, we can adapt the formalism in Sec.~\ref{sec:general_gf} to a DMRG
algorithm by carrying out each step within the renormalized basis at each site.
Again, the simplest procedure to implement is to use a state-averaged
formalism, where all MPS appearing in the equations share
the same renormalized basis $\{ \ket{m[k]_\alpha} \}$ at site $k$.
Thus the four Runge-Kutta vectors in Eq.~\eqref{eq:RK4_vectors} become vectors
in the space of site $k$, $|r_1[k]\rangle \ldots |r_4[k]\rangle$,
and the Hamiltonian used   to construct the vectors is $\hat{H}[k] = P[k] \hat{H} P[k]$.
Note that higher powers of $\hat{H}$ are used in constructing the Runge-Kutta vectors.
Similarly to as in DDMRG, we invoke the approximation:
\begin{align} \label{eq:hpower}
   \hat{H}^n[k] &\approx \hat{H}[k]^n
\end{align}
which again, introduces an error which only vanishes in the limit of infinite bond dimension.
The final consideration is the decimation step to transform from one site to the next.
In td-DMRG, this is done by first computing wavefunctions at the intermediate times
$t+1/3\tau$ and $t+2/3\tau$ using linear combinations
of the $|r[k]\rangle$ vectors:
\begin{align}\label{eq:RK4_interfunc}
&|\Psi(t+\frac{1}{3}\tau)[k]\rangle\approx|\Psi(t)[k]\rangle+\frac{1}{162}[31|r_1[k]\rangle+14|r_2[k]\rangle+14|r_3[k]\rangle-5|r_4[k]\rangle]\nonumber\\
&|\Psi(t+\frac{2}{3}\tau)[k]\rangle\approx|\Psi(t)[k]\rangle+\frac{1}{81}[16|r_1[k]\rangle+20|r_2[k]\rangle+20|r_3[k]\rangle-2|r_4[k]\rangle].
\end{align}
The density matrix used for the renormalization is the weighted average of all the (site) wavefunctions at different times:
\begin{equation}\label{eq:ave_rdm}
\Gamma[k] = w_1\Gamma (|\Psi(t)[k]\rangle ) + w_2\Gamma (|\Psi(t+\frac{1}{3}\tau)[k]\rangle )
          + w_3\Gamma (|\Psi(t+\frac{2}{3}\tau[k]\rangle ) + w_4\Gamma (|\Psi(t+\tau)[k]\rangle ).
\end{equation}
Feiguin and White~\cite{FeiguinPhysRevB2005} found by experimentation that the choice of weights
\begin{equation}\label{eq:weight}
w_1=w_4=\frac{1}{3},\quad w_2=w_3=\frac{1}{6}
\end{equation}
gave the best convergence with bond dimension during the time-propagation.

The accuracy of a td-DMRG simulation is controlled by the bond dimension $M$ as well
as the time-step $\tau$ and total propagation time $T$. In general,
it is found that as $T$ increases, the bond dimension needs to be increased to maintain accuracy
in the wavefunction, due to the generic growth of entanglement during time evolution. Decreasing the time-step decreases the Runge-Kutta integration error, however,
it also increases the number of DMRG sweeps and thus the number of compressions of the wavefunction which
can also lead to an accumulated error.~\cite{FeiguinPhysRevB2005}
Consequently, the time-step should be chosen to balance the intrinsic
time-integration error with the error due to DMRG compressions.

Similarly to DDMRG, the use of a state-averaged renormalized basis at each
site introduces some undesirable errors into the td-DMRG algorithm.
For example, the MPS $\ket{\Psi(t)}$ at the beginning of a time-step,
represented in the renormalized basis at time $t$, becomes approximated
by the renormalized basis at time $t+\tau$ at the end of the time-step, introducing an error
in the representation of the initial state.
Thus, we now consider a more accurate method, where states
at different times are represented by independent MPS. In the most general extension, every state appearing
in the Runge-Kutta scheme would be represented by its own independent MPS,
i.e. $\ket{\Psi(t)}$, $\ket{\Psi(t+\tau)}$, and the Runge-Kutta vectors $\ket{r_1[k]} \ldots \ket{r_4[k]}$.
Operations that increase the bond dimension of the MPS (e.g. when applying the Hamiltonian
to construct the Runge-Kutta vectors, or adding the Runge-Kutta vectors to obtain $\ket{\Psi(t+\tau)})$ are then followed by
variational MPS compression to the desired bond dimension.
We call this scheme, which corresponds to the most direct implementation of time evolution with MPS
in the Runge-Kutta context, td-DMRG(G), to denote the general extension. However, this scheme
is significantly more expensive due to the many compression steps.
A practical compromise is to retain only independent renormalized bases
for $\ket{\Psi(t)}$ and $\ket{\Psi(t+\tau)}$, and to make use of approximations
such as Eq.~(\ref{eq:hpower}) to reduce
the cost. We call this method td-DMRG$^{++}$.
In this case, we construct the four Runge-Kutta states as:
\begin{align}\label{eq:RK4_vec_for_tddmrg++}
&|r_1[k]\rangle = P[k](t+\tau) \tau(\hat{H}-E_0)P[k](t) |\Psi(t)[k]\rangle\nonumber\\
&|r_2[k]\rangle = P[k](t+\tau) \tau(\hat{H}-E_0) P[k](t+\tau) [|\Psi(t)[k]\rangle+1/2|r_1[k]\rangle]\nonumber\\
&|r_3[k]\rangle = P[k](t+\tau) \tau(\hat{H}-E_0) P[k](t+\tau) [|\Psi(t)[k]\rangle+1/2|r_2[k]\rangle]\nonumber\\
&|r_4[k]\rangle = P[k](t+\tau) \tau(\hat{H}-E_0)P[k](t+\tau)  [|\Psi(t)[k]\rangle+|r_3[k]\rangle]
\end{align}
where $P[k](t)$ projects onto the renormalized basis of $\ket{\Psi(t)}$ at site $k$, and $P[k](t+\tau)$ projects
onto the renormalized basis of $\ket{\Psi(t+\tau)}$ at site $k$.
The two sets of renormalized bases $|m[k](t)\rangle$ and $|m[k](t+\tau)\rangle$
are transformed to site $k+1$ using the density matrices of $\ket{\Psi[k](t)}$ and $\ket{\Psi[k](t+\tau)}$
respectively.
More precisely, we use the state average of the density matrices from the real and imaginary parts of the wavefunctions, to ensure
that all tensors in the MPS are real.
Note that if we carried out time-propagation using a first order time-step scheme (involving only the
first Runge-Kutta vector $\ket{r_1[k]}$) then the above procedure is the same as td-DMRG(G), as $P[k](t) | \Psi(t)[k]\rangle$ introduces
no error, and $P[k](t+\tau)$ can viewed as the variational MPS compression (up to the detail of averaging
the real and imaginary wavefunction contributions to the density matrix). At the RK4 level, additional errors
beyond td-DMRG(G) are introduced into the higher Runge-Kutta vectors. However, additional compressions are avoided by
reusing the projected Hamiltonian $\hat{H}[k](t+\tau)$ to construct the additional vectors.
Importantly, the cost of the td-DMRG$^{++}$ method is only a factor of two higher than the standard
td-DMRG procedure of Feiguin and White for blocking and renormalization of the operators,
but as we shall see in the following section, it gives rise to significant improvements in accuracy for a fixed bond dimension,
allowing for time savings in practice.

In summary, the td-DMRG$^{++}$ algorithm consists of:
\begin{itemize}
\item Carrying out ground-state DMRG to obtain $E_0$ and $|\Psi_0\rangle$.
\item Computing the MPS for $a_i^{(\dagger)}|\Psi_0\rangle$.
\item Propagating in real-time for a total time ($T$) as required for the desired accuracy in the
      spectrum. The propagation scheme consists of
      sweeps for each time-step. At each site $k$, we compute
      the four Runge-Kutta vectors using the site Hamiltonians $P[k](t+\tau) \hat{H} P[k](t)$ and $P[k](t+\tau) \hat{H} P[k](t+\tau)$ as
      in Eqs.~\eqref{eq:RK4_vec_for_tddmrg++}.
      We update the renormalized basis for $|\Psi(t+\tau)\rangle$ using the eigenvectors of the density matrix built
      from $|\Psi(t+\tau)\rangle$.
      Sweeps are carried out until convergence in $|\Psi(t+\tau)\rangle$ (typically 2-4 sweeps are sufficient).
\item If desired, $G(t-t')$ is Fourier transformed using Eq.~\eqref{eq:ft} to obtain the frequency-dependent Green's function.
\end{itemize}

\section{Results and Discussion}
\label{sec:results}

\subsection{Benchmarking DDMRG$^{++}$ and td-DMRG$^{++}$}

The DDMRG$^{++}$ and td-DMRG$^{++}$ algorithms above have been implemented inside the \textsc{Block}
DMRG code.~\cite{ChanJChemPhys2002,ChanJChemPhys2004,GhoshJChemPhys2008,SharmaJChemPhys2012}
We now examine the performance of the DDMRG$^{++}$ and td-DMRG$^{++}$ algorithms
in the context of two simple systems where exact results can be computed.
The first is a 10 atom equally spaced hydrogen chain
at the equilibrium bond distance ($r = 1.8$~\bohr (Bohr)) using a minimal STO-6G basis set.\cite{HehreJChemPhys1969}
We shall return to the hydrogen chain problem  in more detail in Section~\ref{subsec:hchains}.
The second is an 8 site 1D Hubbard model with $U = 0.1t$. Except where otherwise stated, we will use spin-adapted
implementations of the algorithms. We found that, similarly to ground-state simulations, spin-adaptation
provides roughly a factor of two gain in the effective bond dimension (see Supplementary Material).

\begin{figure}[!ht]
\includegraphics[width=9cm,trim={1.0cm 0.5cm 2cm 1cm},clip]{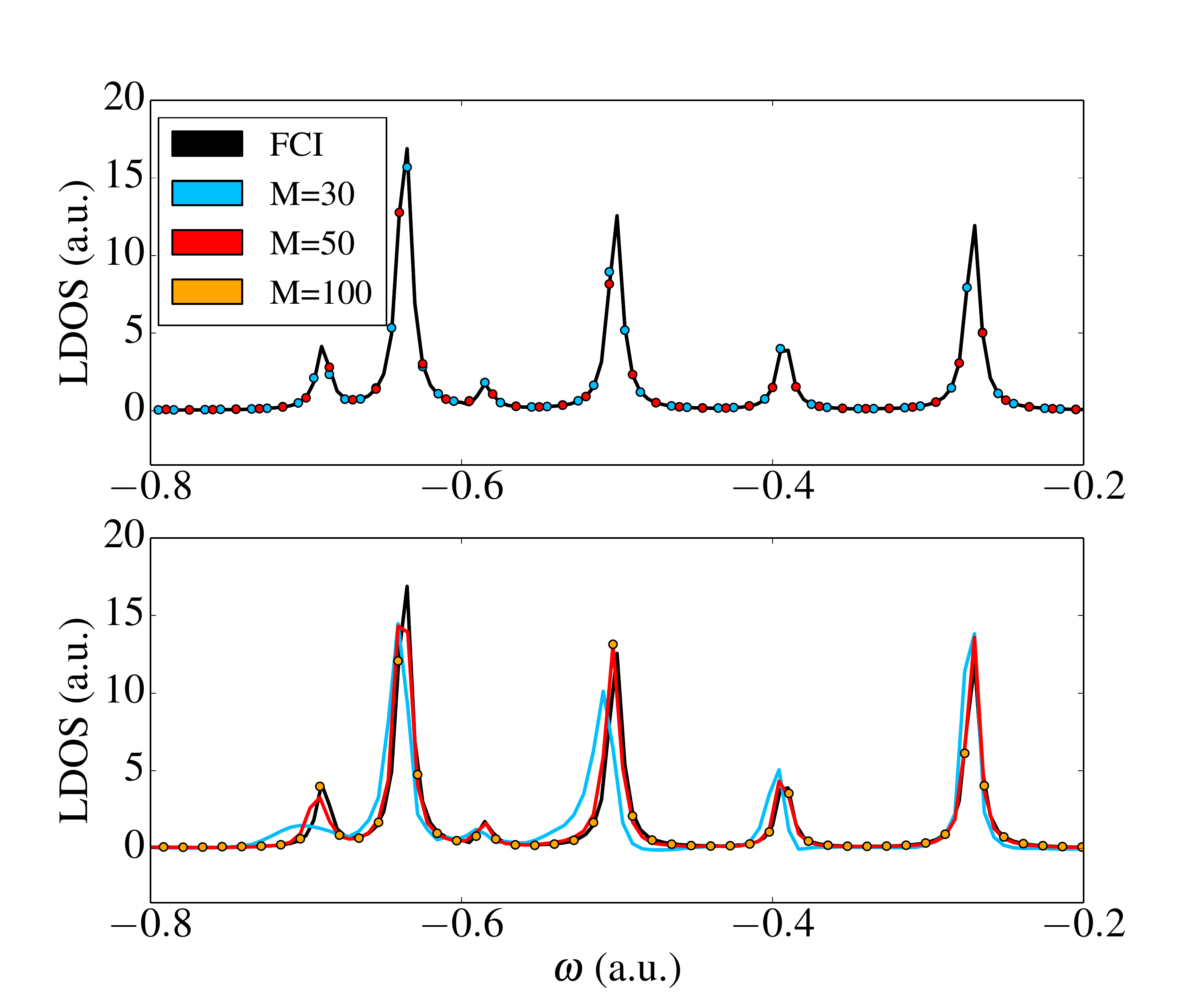}
\caption{\label{fig:h10_td_vs_omega} Dependence of the LDOS on bond dimension $M$ for a \ce{H_{10}} chain
at $r = 1.8$~\bohr.
LDOSs at the central site using DDMRG$^{++}$ (upper panel) and td-DMRG$^{++}$ (lower panel).
A broadening ($\eta$) of 0.005 a.u. has been used. 
For ease of visualization dots and lines are used to represent the same quantity (LDOS); 
different bond dimensions are represented by different colors.
}
\end{figure}

Here we first analyze the performance of DDMRG$^{++}$ and td-DMRG$^{++}$ in the context of the \ce{H_{10}} hydrogen chain. Shown in
Fig.~\ref{fig:h10_td_vs_omega} is the LDOS ($S_{ii}$) ($\eta=0.005$ a.u.) computed with FCI compared against DDMRG$^{++}$
and td-DMRG$^{++}$ ($\tau=0.1$ a.u., $T=1000$ a.u.).
LDOS have been calculated in this case at the central site of the chain starting from 
converged DMRG calculations ($M$=500), and calculations are done in the L{\"o}wdin orthogonalized basis.
To simplify visual comparisons only the IP part of the LDOS is presented here.
From Fig.~\ref{fig:h10_td_vs_omega}, we see that both DDMRG$^{++}$ and td-DMRG$^{++}$ approach
the reference FCI result as $M$ is increased towards the maximum value ($M$=100).
However, DDMRG$^{++}$ converges much more quickly than td-DMRG$^{++}$ toward the exact result.
In particular, DDMRG$^{++}$ is indistinguishable from FCI already at
$M$=30, while td-DMRG$^{++}$ requires $M$=50-100 to reach the same accuracy.
At $M$=30, the td-DMRG$^{++}$ spectrum also has small unphysical negative parts in the
frequency region between -0.5 and -0.3 a.u..
The higher accuracy of the DDMRG$^{++}$ is to be expected given that the algorithm targets a single frequency
at a time.

Analyzing the computational cost of the two algorithms
we have found, for the $\eta$ used,
that the total cost of the DDMRG$^{++}$ and td-DMRG$^{++}$ calculations
(i.e. over all frequencies and for the total propagation time) to reach a given
accuracy is quite similar.
However in many molecular applications, only a small range of frequencies is of interest.
In that case DDMRG$^{++}$ is particularly efficient,
as td-DMRG$^{++}$ computes the spectra over the whole frequency range.
Further, the DDMRG$^{++}$ calculations can be carried
out independently for each frequency point, allowing for easy parallelization.

\begin{figure}[!ht]
\includegraphics[width=9cm,trim={1.0cm 0.0cm 2.0cm 1.cm},clip]{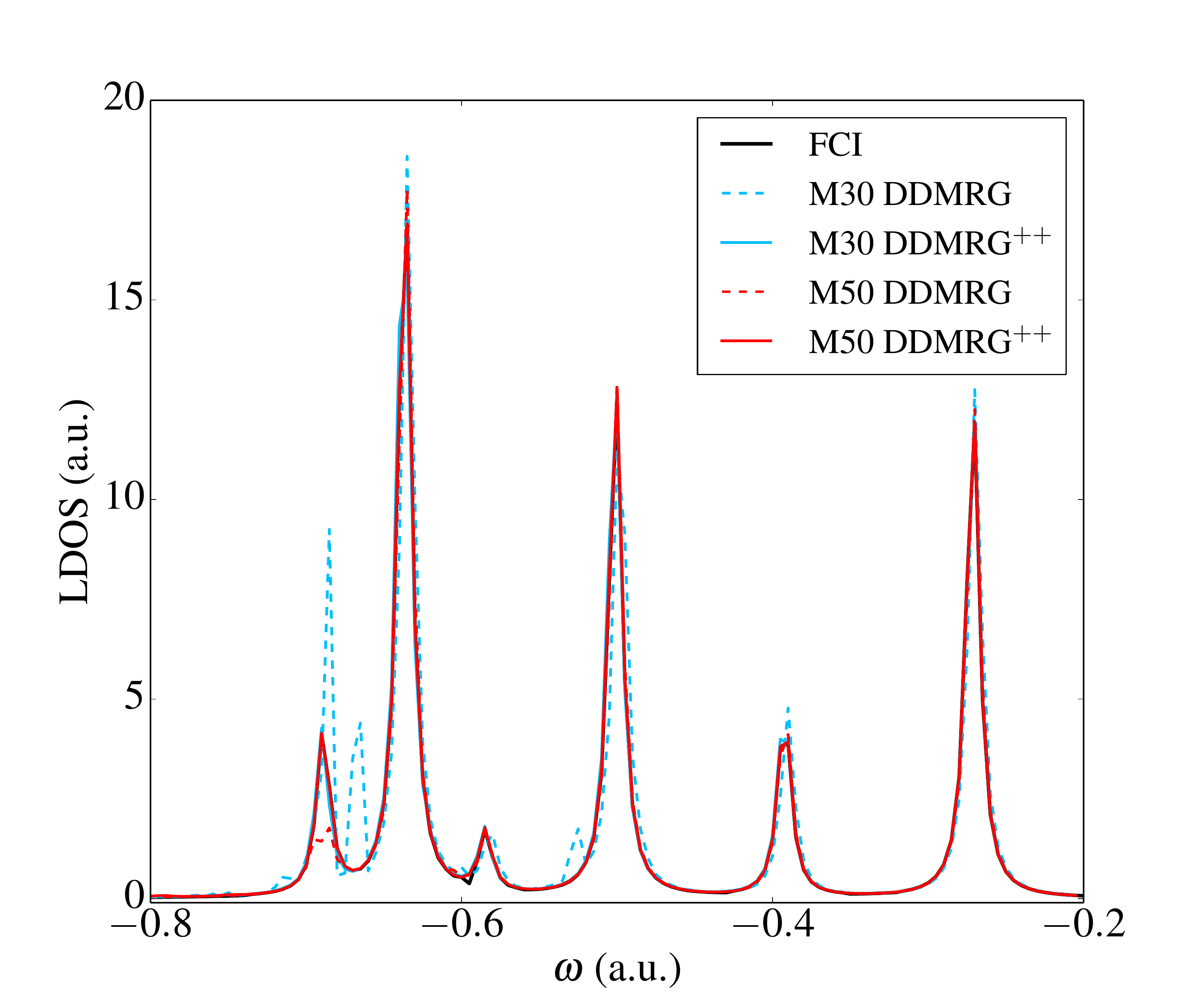}
\caption{\label{fig:h10_ddmrg_compare}
Comparison between DDMRG and  DDMRG$^{++}$ in the description of the spectral function of an equally spaced
10 atom hydrogen chain near the equilibrium bond distance ($r = 1.8$~\bohr).
A broadening ($\eta$) equal to 0.005 a.u. has been used.
}
\end{figure}

Both DDMRG$^{++}$ and td-DMRG$^{++}$ are evolutions of their parent algorithms because they
do not restrict all MPS appearing in the equations to share the same state-averaged basis.
We now examine the effect of this improvement. 
In Fig.~\ref{fig:h10_ddmrg_compare} we compare the DDMRG and DDMRG$^{++}$ algorithms
for the 10 site hydrogen chain. 
While both agree at larger bond dimension (as they must)
for the smaller bond dimension ($M=30$) we see that the DDMRG$^{++}$
spectrum is significantly improved over the DDMRG spectrum, and in particular the DDMRG spectrum
it oscillates, and this is a consequence of representing the ground-state wavefunction
by an MPS in a state-averaged basis with only a small bond dimension. In contrast, even if we
use an $M=30$ ground-state MPS in the DDMRG$^{++}$ algorithm, it has a consistent converged energy across the sweep
which gives rise to a much more stable spectrum.

\begin{figure}[!ht]
\includegraphics[width=9cm,trim={1.0cm 0.0cm 2cm 1cm},clip]{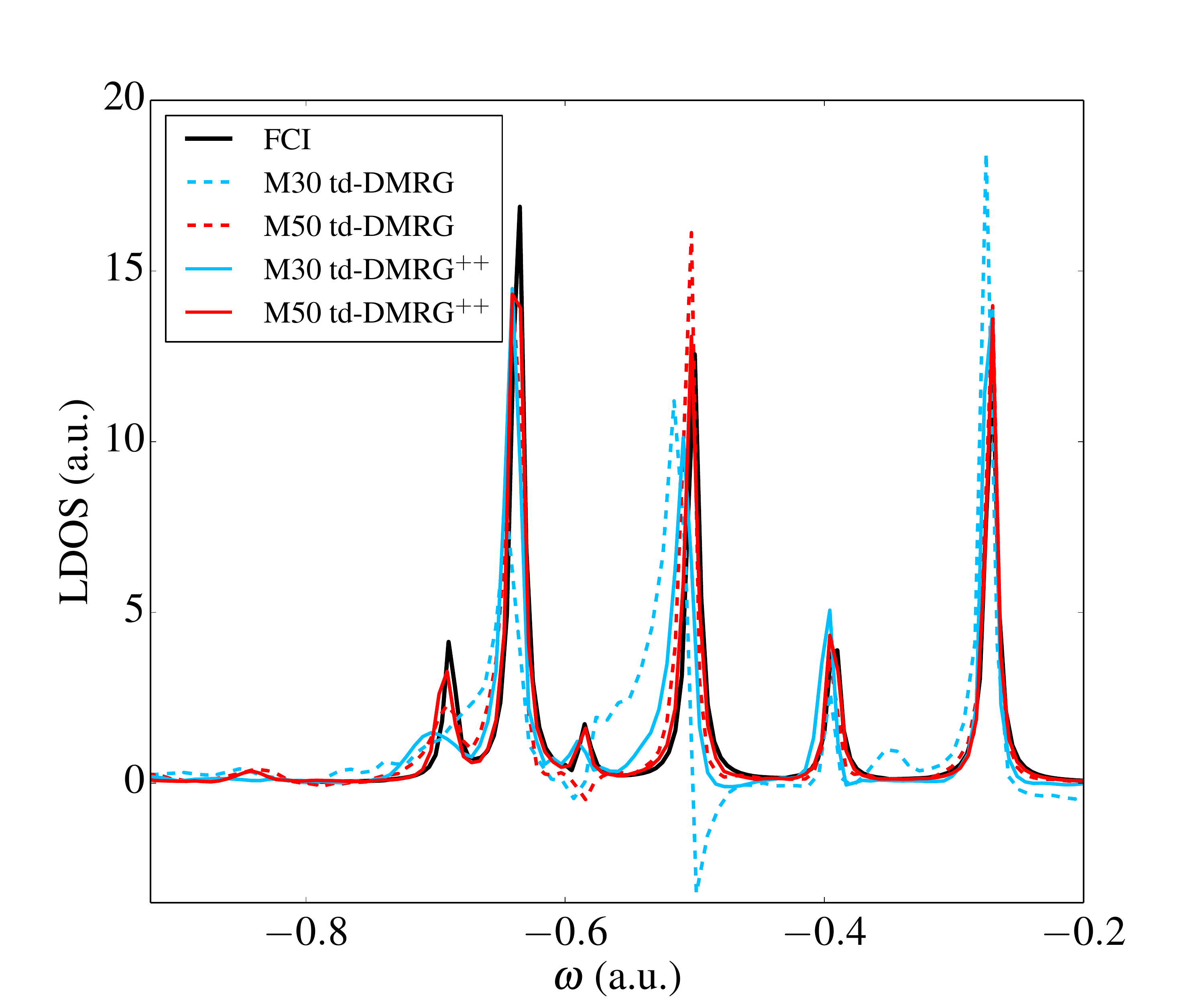}
\caption{\label{fig:h10_rtdmrg}
Comparison between td-DMRG and  td-DMRG$^{++}$ in the description of the spectral function of an equally spaced
10 sites hydrogen chain at the equilibrium bond distance ($r = 1.8$~\bohr).
A broadening ($\eta$) equal to 0.005 a.u. has been used.
}
\end{figure}
In Fig.~\ref{fig:h10_rtdmrg} we compare the td-DMRG and td-DMRG$^{++}$ algorithms for
the 10 site hydrogen chain (\ce{H_{10}}). We see that the $M=30$ td-DMRG$^{++}$
calculation is comparable in accuracy, if not better than, the $M=50$ td-DMRG calculation. 

In both the DDMRG$^{++}$ and td-DMRG$^{++}$ cases, the cost of the calculations for fixed bond dimension
is roughly twice the cost of the original DDMRG and td-DMRG algorithm. On the other
hand, the effective bond dimension in DDMRG$^{++}$ and td-DMRG$^{++}$ appears to be close to twice
that in DDMRG and td-DMRG respectively. Given that the scaling with bond dimension
is like $M^3$, we see that the DDMRG$^{++}$ and td-DMRG$^{++}$ algorithms offer significant savings in practice.

\begin{figure*}[!ht]
\centering
\includegraphics[width=9cm,trim={9cm 2cm 22cm 0cm}]{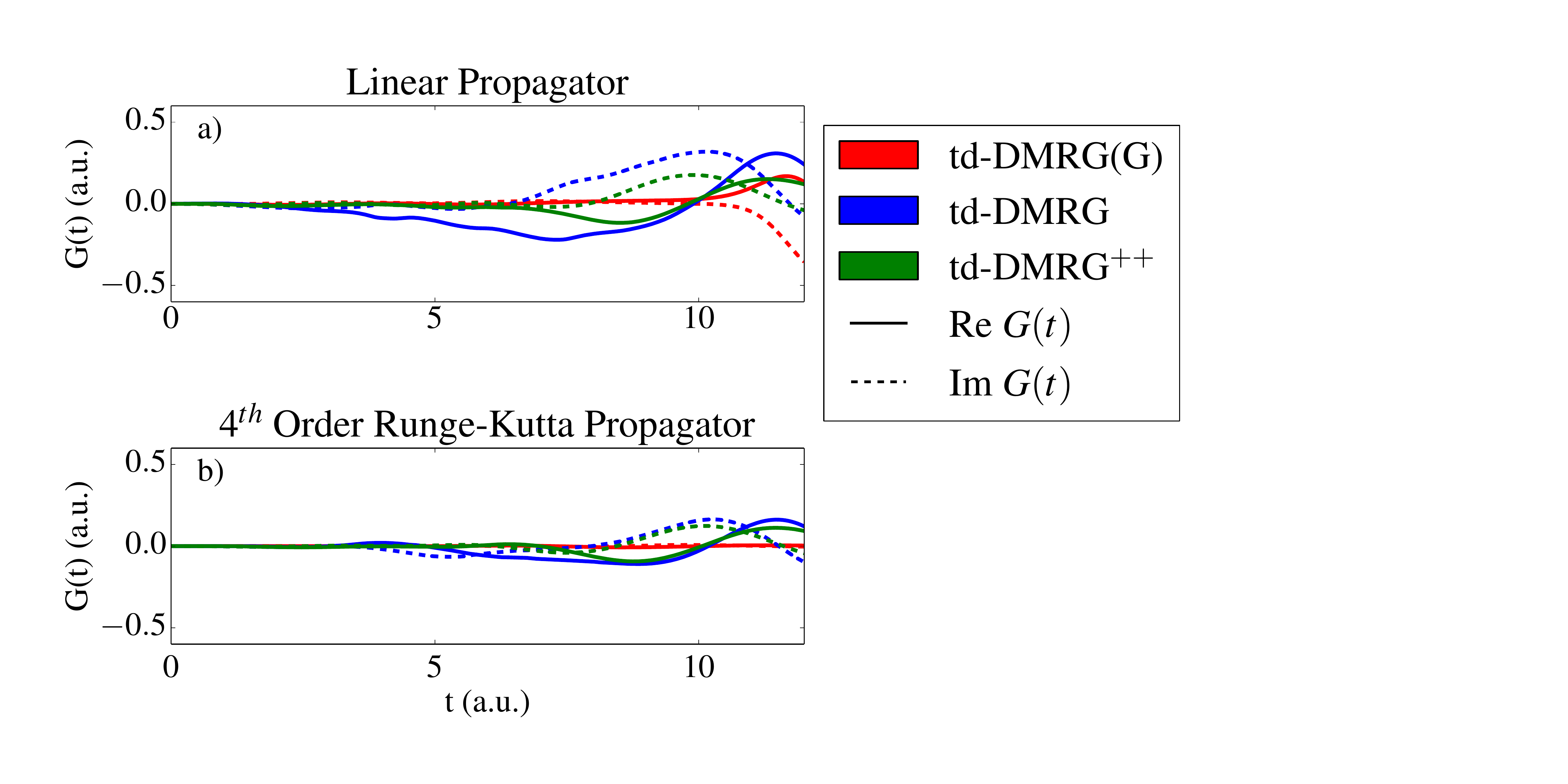}
\caption{\label{fig:Hub_MPO_time_diffmet} Errors of td-DMRG, td-DMRG$^{++}$ and td-DMRG(G) in the estimation
of the $G_{00}(t)$ matrix element of a 8-site Hubbard chain with respect to the exact FCI propagation.
Results obtained using the linear propagator and the 4th order Runge-Kutta (RK4) scheme are presented in panel
$a$ and $b$ respectively.
In this plot different colors refer to different methods and solid and dashed lines are used to 
represent the real and imaginary parts of $G_{00}(t)$ respectively.
}
\end{figure*}

Additional understanding of the behaviour of td-DMRG$^{++}$ can be obtained comparing the time-dependent Green's function matrix elements ($G_{00}(t)$ in this case) calculated
with td-DMRG, td-DMRG$^{++}$, and td-DMRG(G) using $M=30$ with both a linear propagator, and the
4th order Runge-Kutta propagator.
Because of the cost of the td-DMRG(G) algorithm, which requires variational MPS compression at each time step, we
performed comparisons for the simpler case of the 8-site Hubbard chain.
Plots of the errors calculated against the exact FCI propagation are presented in Fig.~\ref{fig:Hub_MPO_time_diffmet}.
The td-DMRG(G) calculations were carried out with a general purpose MPO/MPS library without
spin-adaptation~\cite{LiJChemTheoryComput2017}, and thus all calculations in the figure did not use spin adaptation.
We see that both with the linear propagator and the 4th order propagator, the use of more flexible renormalized bases in td-DMRG$^{++}$
and td-DMRG(G) significantly increases the accuracy of the propagation over the simple td-DMRG scheme. In particular, td-DMRG$^{++}$
roughly allows for a doubling of the propagation time over td-DMRG before a noticeable buildup of error occurs, while
td-DMRG(G) allows for a further doubling.
In the case of the linear propagator, the only difference in principle between td-DMRG$^{++}$ and td-DMRG(G) is the use
of the real and imaginary averaged density matrix to determine the renormalized bases (compression) for the wavefunction
at the next time-step (first case), rather than the exact MPS variational compression algorithm (latter method),\footnote{A single complex density matrix is
used in td-DMRG(G) during the compression step.} and this is responsible for
the difference in accuracy.
In the case of the 4th order propagation scheme, td-DMRG(G) provides an accurate representation
of all the Runge-Kutta vectors. This leads to an extremely stable propagation,
but at the cost of a significantly larger number of compression steps (6 more compressions per time step).
Based on this analysis, we can conclude that td-DMRG$^{++}$ provides a good compromise between
accuracy in the representation of the Runge-Kutta vectors, and efficiency in practice, when carrying out real-time propagation.
Note that the error due to finite $T$ is smaller than the other
errors analyzed in this section and thus we have not discussed it in detail. 
A more detailed analysis of the errors associated with the time-step $\tau$ is presented in the supplementary material.

\subsection{Core-ionization potential of \ce{H_2O}}
\label{subsec:h2o}

As a chemical application of the methods developed here
we now consider the calculation of a core-ionization potential.
Core spectra are generally challenging to simulate as they need
a flexible treatment of electron correlation
as well as the inclusion of relativistic effects~\cite{CorianiJChemPhys2015,
DuttaJChemTheoryComput2015,WenzelJComputChem2015,BradecJChemPhys2012}.
Here, we use DDMRG$^{++}$ to calculate
the ionization potential (IP) for the deepest core orbital (O $1s$) of water examining
the basis set effects and the effects of relativity.
We compare against coupled cluster calculations~\cite{DuttaJChemTheoryComput2015,CorianiJChemPhys2015},
as well as experimental reference values in Table~\ref{tab:h2o_ip}.
\begin{table*}[!ht]
\caption{\label{tab:h2o_ip} \ce{H_2O} core ionization potentials (eV).
Theoretical data have been calculated at the geometry of Ref.~\citenum{SenMolPhys2013}.}
\begin{threeparttable}
{\footnotesize
\begin{tabular}{l@{\hskip .2in}c@{\hskip .2in}c@{\hskip .2in}c@{\hskip .2in}c@{\hskip .2in}c@{\hskip .2in}r}
\hline
\hline
                             &CVS-           &EOM-            &EOM-                 &$\Delta$UGA-    &               &                \\
Basis                        &CCSD\tnote{a}  &CCSD            &CCSD(2)$^*$\tnote{c} &SUMRCC\tnote{b} &DDMRG$^{++}$  &Exp.\tnote{d}   \\
\hline
cc-pVDZ                      &543.34         &543.27\tnote{b} &                     &541.97          &542.13         &539.78          \\
cc-pVTZ                      &540.68         &540.66\tnote{b} &                     &539.02          &539.62         &                \\
cc-pVQZ                      &               &                &                     &                &539.73         &                \\
cc-pCVDZ                     &               &542.69\tnote{c} &541.17               &                &541.30         &                \\
cc-pCVTZ                     &541.15         &541.13\tnote{c} &540.03               &                &540.10         &                \\
cc-pVDZ-DK\tnote{$\Diamond$} &               &                &                     &                &542.53         &                \\
cc-pVTZ-DK\tnote{$\Diamond$} &               &                &                     &                &539.96         &                \\
cc-pVQZ-DK\tnote{$\Diamond$} &               &                &                     &                &540.16         &                \\
\hline
\hline
\end{tabular}
\begin{tablenotes}
\item[$\Diamond$] Scalar relativistic effects have been introduced using the sf-X2C method.\cite{LiuMolPhys2010,SaueChemPhysChem2011,PengTheoChemAcc2012}
\item[a] Data from Ref.~\citenum{CorianiJChemPhys2016,CorianiJChemPhys2015}
\item[b] Data from Ref.~\citenum{SenMolPhys2013}
\item[c] Data from Ref.~\citenum{DuttaJChemTheoryComput2015}
\item[d] Data from Ref.~\citenum{OhtsukaJChemPhys2006}
\end{tablenotes}
}
\end{threeparttable}
\end{table*}
We estimate the IP from a DDMRG$^{++}$ calculation by fitting three points around the excitation peak
with a parabola and extracting the position of the maximum. We used an $\omega$ grid of 0.01 hartree
and an $\eta$ value of 0.05 hartree.
We used a bond dimension large enough to converge the DMRG energy below the milliHartree (\mEh) level
($M$=1000 for DZ basis sets and $M$=2000 for TZ and QZ basis sets),
while a bond dimension $M$=500 has been used in DDMRG$^{++}$ to represent the
$a_i|\psi_0\rangle$ and $|c(\omega)\rangle$ wave functions.
Calculations using smaller bond dimensions in the cc-pVQZ basis indicate that our IP results are converged
to better than 0.1 eV. Smaller errors are expected for the smaller basis sets.

Overall, our computed IP's are in general agreement with previous theoretical results and, if we use a basis set
larger than double zeta (DZ), they are in good agreement with the experimental value as well.
As noted above, relativistic effects are important for this quantity.
Four component relativistic DMRG calculations have previously been reported in {Ref.~\citenum{KnechtJChemPhys2014}};
here we estimate scalar relativistic corrections through the sf-X2C Hamiltonian.~\cite{LiuMolPhys2010,SaueChemPhysChem2011,PengTheoChemAcc2012}  
The inclusion of scalar relativistic effects increases the IP by 0.35-0.4 eV. The final result in the largest cc-pVQZ basis including
scalar relativistic effects is within 0.4 eV of the experimental value.
The core-valence basis sets shift the ionization potential by a similar amount but with a different sign at the DZ and TZ level.

The DDMRG$^{++}$ calculations allow for an assessment of correlation effects beyond those treated in earlier methods. Comparing
to the EOM-CCSD and CVS-CCSD results, we find that the correlation effects beyond doubles amount to approximately 1 eV in the IP. 
Interestingly, the EOM-CCSD(2)* method recently developed by Dutta et al\cite{DuttaJChemTheoryComput2015} performs very well, with errors of roughly 0.1 eV. 
MRCC ($\Delta$UGA-SUMRCC) calculations, as performed by Sen et al in Ref.~\citenum{SenMolPhys2013} also improve on the EOM-CCSD
results. 

\subsection{Hydrogen Chains}
\label{subsec:hchains}

We now use the methods developed in this work to study longer hydrogen chains.
1D equally spaced hydrogen chains were introduced in Ref.~\citenum{HachmannJChemPhys2006} as a simple model for strong correlation
in an ab initio system, with the tuning parameter being the spacing between the atoms (here denoted $r$). They have since
become a popular model system on which to benchmark strong correlation methods~\cite{StellaPhysRevB2011,TsuchimochiJChemPhys2009,
  LinPhysRevLett2011,SinitskiyJChemPhys2010,MazziottiPhysRevLett2011,MottaHChain1ArXiv2017}, and have also spawned the study
of analogous ring systems with heavier atoms~\cite{FertittaPhysRevB2014,wouters2016practical}. In the thermodynamic limit, the chains
are thought to undergo a metal-insulator transition with the metallic phase being found
at short bond distances and a Mott insulator found at long distances. 
1D hydrogen chains also serve as a dimensionally reduced setting to study 
the hydrogen phase diagram,  which is of particular interest in understanding 
the high pressure interiors of planets such as Jupiter and Saturn.

The metal-insulator transition in hydrogen chains can be identified in terms of different observables.
Direct evidence can be obtained by computing the bandgap in the thermodynamic limit, which must vanish for a metal.
Alternatively, ground-state correlation functions can be computed. 
For a 1D system, the delocalization of the electrons associated with the metallic phase can be established by the vanishing of 
the many-body complex polarization function~\cite{RestaRevModPhys1994,RestaPhysRevLett1998,StellaPhysRevB2011,HineJPhysCondensedMatter2007}.
Also, the algebraic decay of the off-diagonal elements of the single-particle density matrix can also be used to
establish the metallic phase~\cite{HachmannJChemPhys2006}. This latter criterion
was used in earlier DMRG studies to characterize the metallicity
of hydrogen chains at different bond lengths~{\cite{HachmannJChemPhys2006}.

Here we use the DDMRG$^{++}$ and td-DMRG$^{++}$ algorithms to calculate
the LDOS and the complex polarization function respectively as measures of metallicity, as a function
of bond length for three different hydrogen chains in the minimal STO-6G basis set~\cite{HehreJChemPhys1969} with
open (OBC) and periodic boundary conditions (PBC).
We also carry out ground-state DMRG and restricted and unrestricted Hartree-Fock calculations to
further support the results. All DMRG calculations are carried out with localized 
L\"owdin orthogonalized atomic orbitals, and LDOS are presented at one of the (two) central atoms of the chain.
The PBC Hamiltonian is defined using a periodic Coulomb interaction only along the chain (1D periodicity).

In Fig.~\ref{fig:hchains_diffdist} we present the DDMRG$^{++}$ LDOS at three bond distances,
$r=1.4, 1.8, 3.6\, \bohr$ for 10, 30, and 50 atom hydrogen chains using open boundary conditions.
For these systems $r=1.8$~\bohr is close to the equilibrium bond distance.\cite{HachmannJChemPhys2006,MottaHChain1ArXiv2017}
The PBC spectral functions for \ce{H_{50}} at two different geometries ($r=1.4, 3.6\, \bohr$) are also shown.
Additional OBC LDOS e.g. for intermediate bond distance can be found in the Supplementary Material.
\begin{figure}[!ht]
\includegraphics[width=9cm,trim={1.0cm 0.0cm 2.2cm 1.5cm},clip]{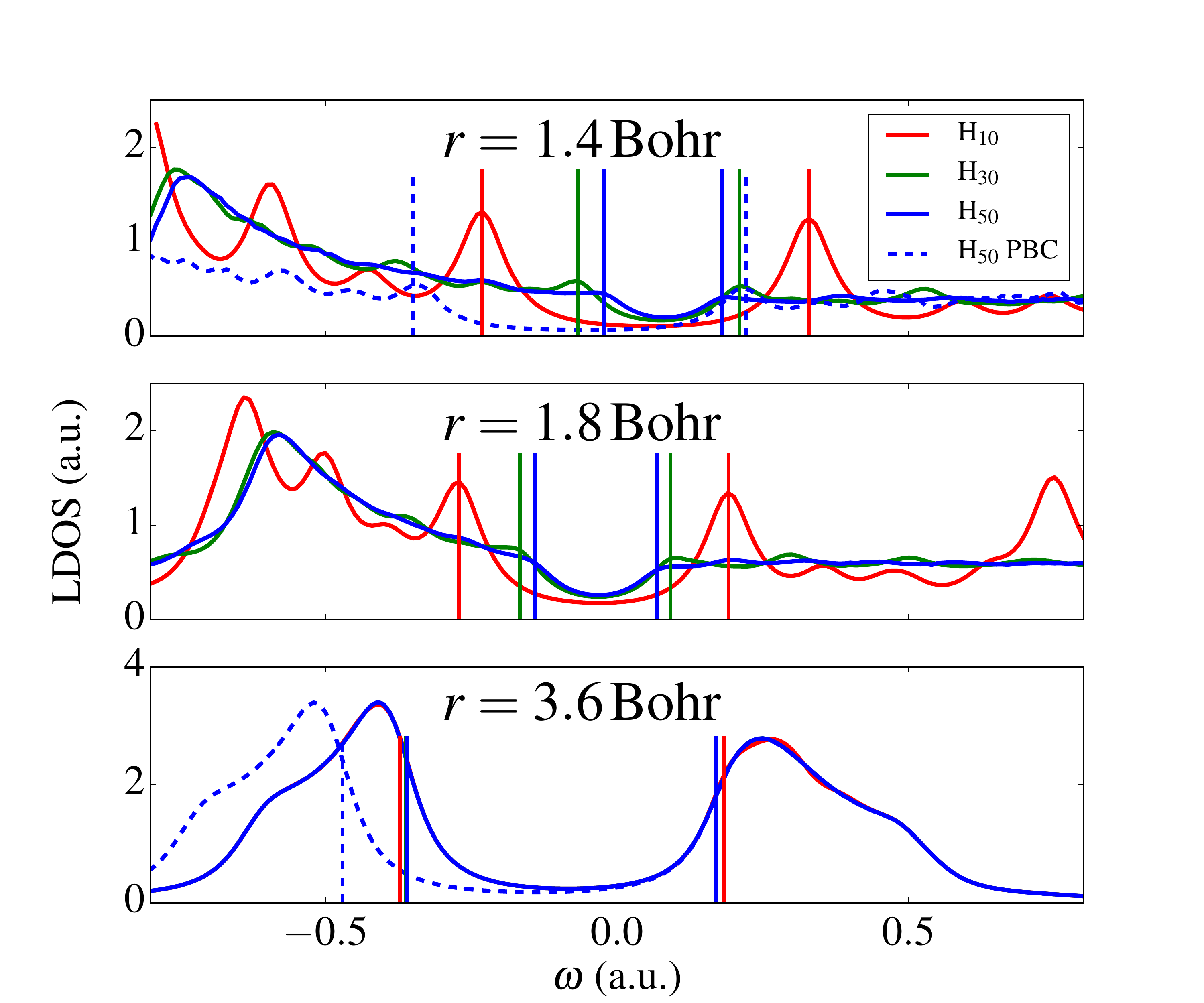}
\caption{\label{fig:hchains_diffdist} LDOS for three equally spaced hydrogen chains
(\ce{H_{10}} - red, \ce{H_{30}} - green, \ce{H_{50}} - blue) at three different bond distances ($r = 1.4, 1.8$ and $3.6$~\bohr).
LDOS of \ce{H_{50}} calculated at bond distances $r = 1.4$ and $3.6$~\bohr with PBC (blue dashed lines)
have been also included.
All the LDOSs have been calculated on the central site of the chain.
A broadening ($\eta$) of 0.05 a.u. has been used.
Vertical lines are used to indicate the position of the ionization potential (IP) and
electron affinity (EA) of the systems.}
\end{figure}

As the chain-length is increased, the gap is reduced but does not yet close. The finite size effects
for \ce{H_{50}} at $r=1.8\, \bohr$ and $r=3.6\, \bohr$ are well converged as one can observe by comparing
the \ce{H_{30}} and \ce{H_{50}}  LDOS.
However, significant finite size effects start appearing for more compressed chains, as can be seen
for the $r=1.4\, \bohr$ chain. We note that for compressed chains, the finite size
error is mainly a single-particle effect rather than a result of Coulomb interactions. This is because
the kinetic energy scales as $1/r^2$ at small $r$ while the Coulomb energy scales as $1/r$.

The DDMRG$^{++}$ bandgap decreases significantly as the bondlength is decreased from 3.6 to 1.4~\bohr.
As the broadening in the LDOS blurs the gap, it is difficult to determine the gap with high precision
purely from the LDOS.
For this reason we also show positions of the ionization potential 
(IP) and electron affinity (EA) (vertical lines) computed from ground-state DMRG calculations 
at the same geometries. 
Note that (up to finite bond dimension errors) these will occur at precisely the same position 
as the rightmost and leftmost energy poles of the IP and EA Green's function computed from DDMRG$^{++}$.
Determining the gap from IP-EA for the \ce{H_{50}} chain, for instance, gives 202, 209, 
and 530~\mEh gaps for  $r=1.4, 1.8, 3.6$~\bohr respectively. 

The finite chain gaps with OBC and PBC are not entirely consistent, and unfortunately it is difficult to estimate the band gaps in the thermodynamic limit.
With PBC in particular, there are spurious interactions between charges and the periodic images
of their exchange-correlation holes, and this produces larger finite size effects in the PBC calculations than in the OBC
calculations, leading to a very poor thermodynamic limit extrapolation with PBC. Note that both the finite chain OBC and PBC gaps start to
{\it increase} at very compressed distances due to the large single-particle finite size effects discussed above.

\begin{figure}[!ht]
\includegraphics[width=9cm,trim={1.0cm 0.0cm 2.2cm 1.5cm},clip]{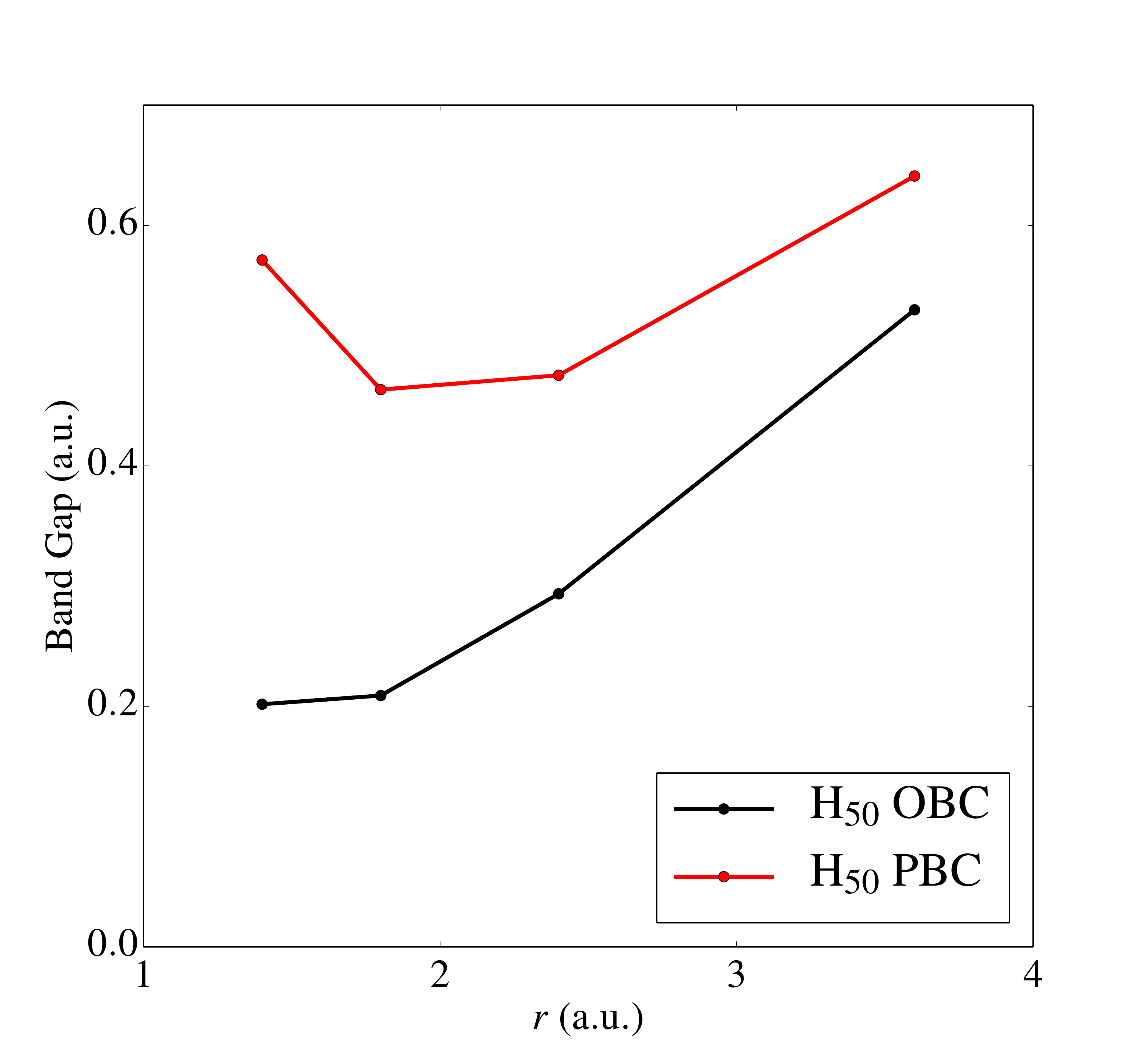}
\caption{\label{fig:GAP_vs_R}
Band gaps calculated for the \ce{H_{50}} chain. 
}
\end{figure}

To understand the effect of correlation on the metallicity, we show for comparison the RHF and UHF results. 
Both the RHF and UHF solutions display gaps, and at short distances, the RHF gap agrees well
with the DMRG gap; the RHF gap at
$r=1.4$~\bohr for \ce{H_{50}}, for instance, is very similar to the DMRG reference (RHF = 234~\mEh, DMRG = 202~\mEh). 
At longer distances, the RHF gap is too small and is only 175~\mEh at $r=3.6$~\bohr while the DMRG gap is 530~\mEh. 
The behaviour of the UHF gap with bond distance is qualitatively correct, but
UHF overestimates the gap at all distances (e.g. for \ce{H_{50}} at $r=1.4$~\bohr it is 312~\mEh while at $r=3.6$~\bohr it is 734~\mEh). 
Note that at longer distances, the RHF gap is not a simple finite size effect but arises from the dimerization of the RHF solution 
through a bond-order wave, as can be clearly seen from the off-diagonal 
bond-order matrix elements of the 1-particle density matrix 
(i.e. $\rho_{i,i+1}$, $\rho_{i+1,i+2}$) see Fig.~\ref{fig:h50_density_trends}.
The DMRG gaps are bounded by the RHF and UHF gaps for $r> 1.8$~\bohr.
\begin{figure}[!ht]
\includegraphics[width=9cm,trim={1.0cm 0.0cm 2.2cm 1.5cm},clip]{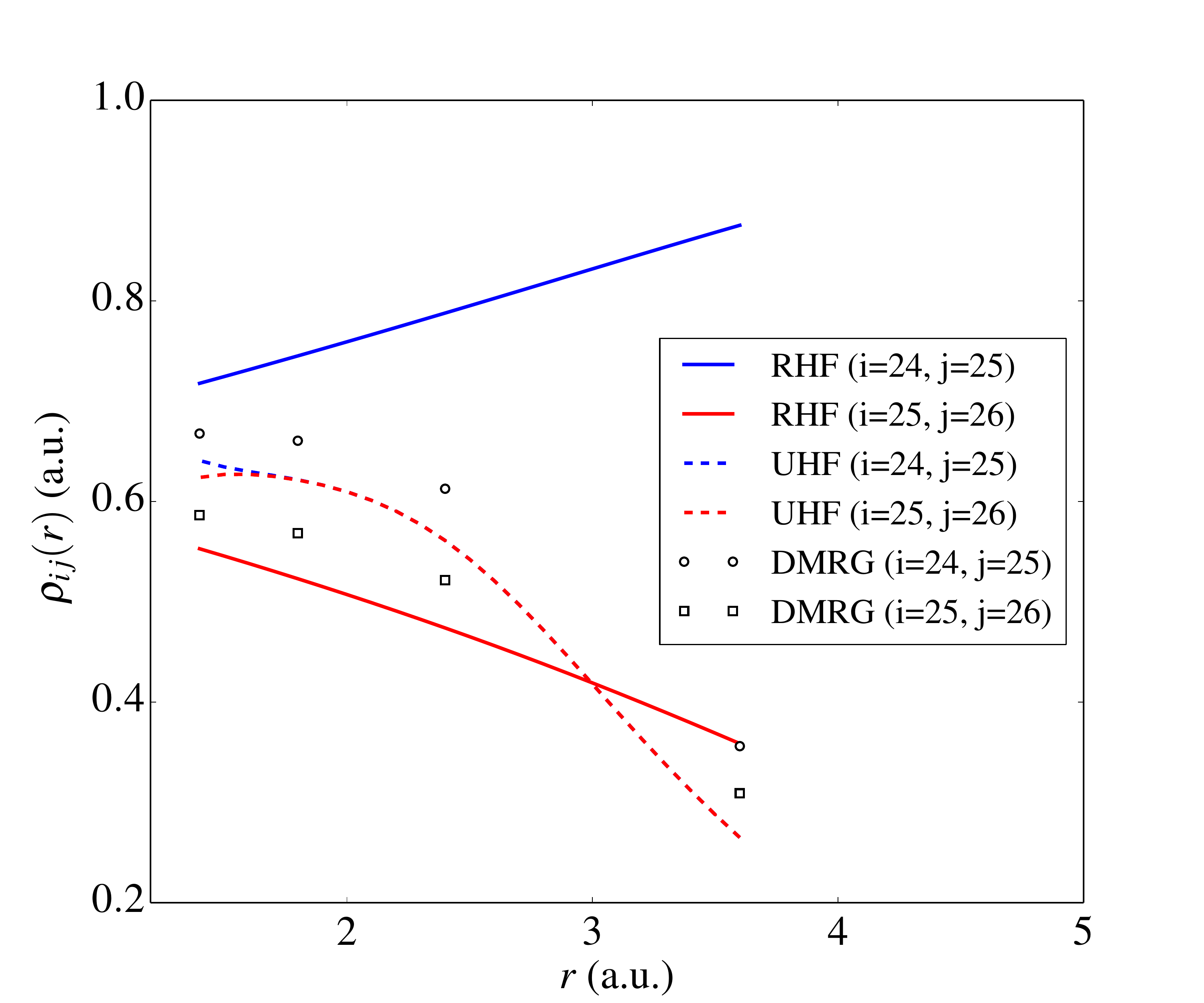}
\caption{\label{fig:h50_density_trends}
Comparison of DDMRG$^{++}$ (dots), RHF (solid line) and UHF (dashed line)
density matrix off-diagonal elements $\rho_{ij}$ for the equally spaced \ce{H_{50}} chain as a function of the
 bond distance.}
\end{figure}

Another way to characterize the metallicity of the ground-state is from
the complex polarization function. This quantity, denoted 
$\tilde{z}$,\cite{RestaPhysRevLett1998,StellaPhysRevB2011} is defined as:
\begin{equation}\label{eq:comppol}
\tilde{z}=\langle\Psi_0|e^{i(2\pi/N)\sum_ir_{in}}|\Psi_0\rangle
\end{equation}
where $r_{in}$ is the component of the $i^{th}$ electron position vector
along the chain axis ($z$ in this case) and $N$ is the longitudinal dimension
of the supercell. 
$\tilde{z}$ measures electron delocalization in the ground-state and its
modulus $|\tilde{z}|\rightarrow 0$ for metallic behaviour, 
while $|\tilde{z}|\rightarrow 1$ in an insulator.
Although $\tilde{z}$ is a complicated many-body observable, it can be conveniently computed by carrying
out a time evolution for unit time using the fictitious Hamiltonian $\hat{H} = 2\pi/N\sum_ir_{in}$, followed
by evaluating the overlap with the ground-state. Here we compute  $\tilde{z}$ using the td-DMRG$^{++}$ algorithm.
Note that when PBC are imposed the direct calculation of dipole integrals is not possible.\cite{RestaPhysRevLett1998}
Given the local character of the Gaussian basis used, we define
the dipole integrals as a multiplicative operator over the basis functions of the reference cell, such that: 
$\langle k |r|l\rangle \approx i\delta_{kl}$ where $i$ is the dimensionless number that indexes the
position of the site $i$ on the chain. In the metallic limit, where the wavefunction is a product state of Bloch functions
built from a single atom unit cell, this approximation yields $\tilde{z}=0$ as an exact evaluation would,
and further the approximation becomes exact in the limit of long bond distances.
\begin{figure*}[!ht]
\includegraphics[width=16cm,trim={0.0cm 5.0cm 0.0cm 5.0cm},clip]{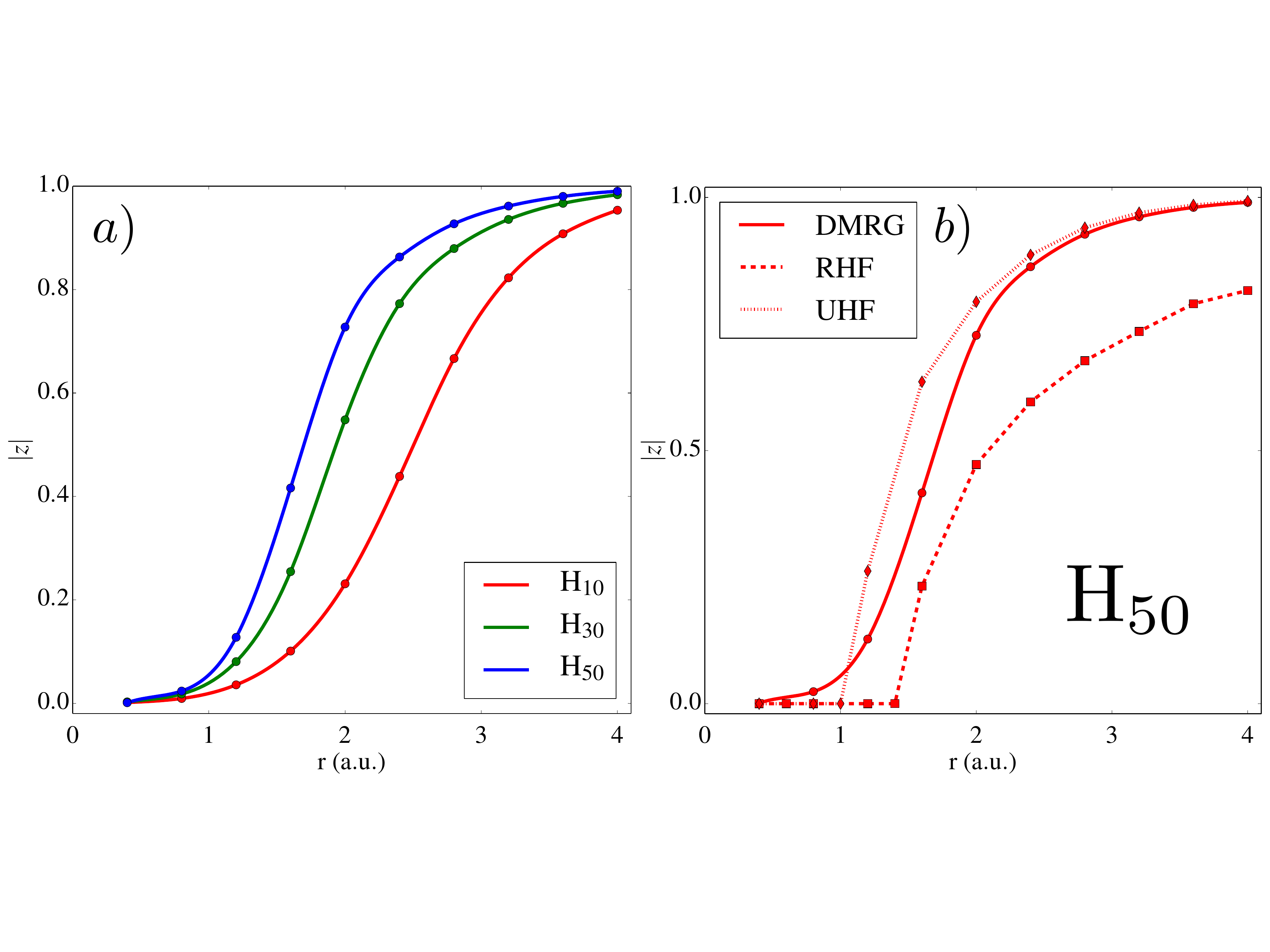}
\caption{\label{fig:comppol}
DMRG and HF complex polarization functions.
In panel a) complex polarization functions  for \ce{H_{10}}, \ce{H_{30}} and \ce{H_{50}} 
using DMRG are presented.
In panel b) complex polarization functions  for \ce{H_{50}} at the DMRG, RHF and UHF 
level of theory are presented. Periodic Boundary Conditions (PBC) have been used each case. 
}
\end{figure*}

In Fig.~\ref{fig:comppol} we plot the DMRG complex polarization function for \ce{H_{10}}, \ce{H_{30}}, and \ce{H_{50}} with PBC;
for the \ce{H_{50}} chain we compare with the RHF and UHF values. The absolute value of the complex polarization function is
exponentially sensitive to localization length and 
decreases very rapidly, for \ce{H_{50}} for instance, near $r=2.0$~\bohr, and becomes close to zero for $r< 1.0$~\bohr.
A similar picture is presented by the RHF and UHF complex polarization functions.
Unlike the single-particle gap, the complex polarization function can vanish in a system even when single-particle
finite size effects are large so long as the electrons are completely
delocalized. The vanishing of the DMRG complex polarization function in this system at short distances,
as also reflected by the similarity in the size of the gaps, thus reflects
the fact that the DMRG wavefunction begins to resemble the RHF wavefunction which is a Slater determinant of plane-wave
like orbitals. However, the scaling of the complex polarization function with system size, much like the gap, converges
only slowly with system size. Thus, to definitively establish a metal insulator transition will require studies of larger systems.
These studies will be discussed in a future publication.

\section{Conclusions}
\label{sec:conclusions}

In this work we studied two algorithms to obtain dynamical quantities from density matrix renormalization group wavefunctions in the ab initio context: the dynamical DMRG (DDMRG) algorithm,
and the time-step targeting time-dependent DMRG (td-DMRG) algorithm. In particular, we proposed and implemented two improved variants of these algorithms, DDMRG$^{++}$ and td-DMRG$^{++}$, in the 
context of computing Green's functions and the density of states. DDMRG$^{++}$ and td-DMRG$^{++}$ yield  improved dynamical quantities with respect to their parent DDMRG and td-DMRG algorithms, at a nominal increase in cost,
and they are both simple to implement within existing ab initio DMRG codes.
Our analysis suggests that DDMRG$^{++}$ and td-DMRG$^{++}$ require a comparable amount of computation time if we desire the density of states at a similar resolution over a large energy window. However,
if one is interested only in the density of states in a small energy window (e.g. when computing the principal core ionization peak) then DDMRG$^{++}$ is advantageous.

In our applications, we showed that in the water molecule, we could use DDMRG$^{++}$ to compute a core excitation energy in a quadruple zeta basis
at a benchmark level of quality beyond that of existing correlation treatments. This suggests that DDMRG$^{++}$ and td-DMRG$^{++}$ will 
provide benchmarking capabilities for ab initio dynamical quantities similar to that provided by ground-state DMRG for ground-state properties.
We also showed in larger hydrogen chains that we could use DDMRG$^{++}$ to compute the ab initio density of states in a system large enough to consider the thermodynamic limit of the spectrum, and
used td-DMRG$^{++}$ to compute a complicated measure of delocalization, the complex polarization function. Both these capabilities will be useful in establishing the physics of the correlated metal-insulator transition in
hydrogen chains, and more broadly to approach the spectral functions of other complex condensed phase problems in the future. Finally, the feasibility of these calculations 
suggests that DDMRG$^{++}$ and td-DMRG$^{++}$ may be fruitfully used to study the correlated density of states of more complex chemical
systems, such as
the multi-centre transition metal complexes that have previously been studied with DMRG. These are directions we will pursue in the future.

\begin{acknowledgement}
This work was supported by the US National Science Foundation via NSF:CHE-1657286 and NSF:CHE-1650436. E.R. would like to thank
Dr. Alexander Yu. Sokolov for insightful discussions and Dr. Weifeng Hu for his help
with the \textsc{Block} DMRG code.
\end{acknowledgement}


\bibliography{dmrg_gf}
\clearpage

\begin{suppinfo}
Supplementary materials: effects of spin-adaptation and time-step size 
on the accuracy of td-DMRG/td-DMRG$^{++}$,  additional spectral functions of hydrogen chains.
\end{suppinfo}
\clearpage



\section*{Supplementary material}

# Supplementary Material: Time-step targeting time-dependent and dynamical density matrix renormalization group algorithms with ab initio Hamiltonians

Enrico Ronca,[*] Zhendong Li, Carlos A. Jimenez-Hoyos, and

Garnet Kin-Lic Chan[*]

*Division of Chemistry and Chemical Engineering, California Institute of Technology,*

*Pasadena, CA 91125, USA*

E-mail: enrico.r8729@gmail.com; gkc1000@gmail.com

# 1 Spin-Adaptation effects on td-DMRG LDOS

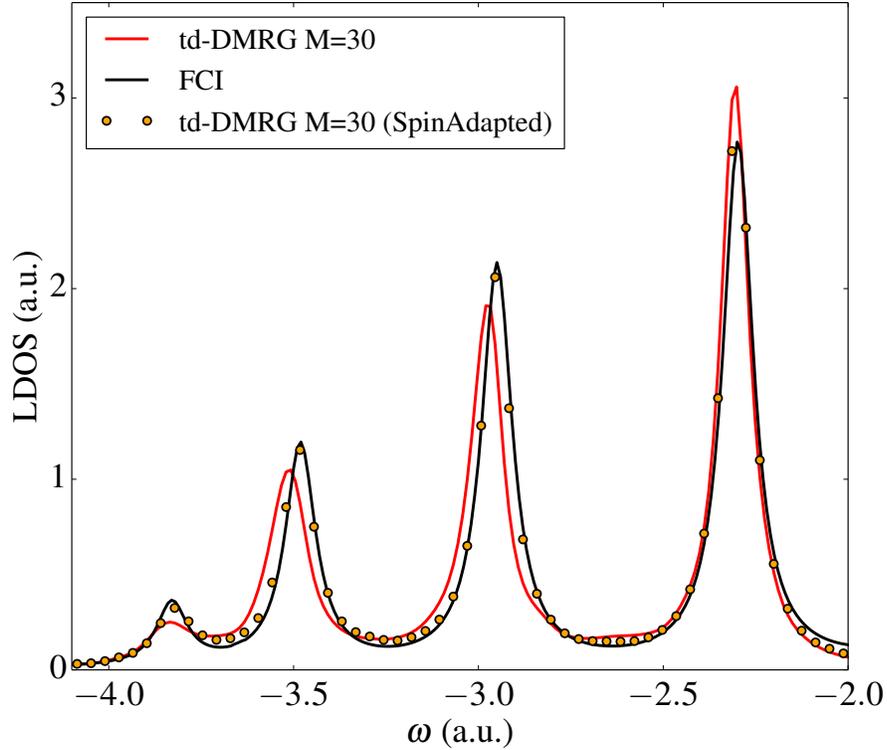

Figure 1: Effect of Spin-Adaptation on the Spectral Function of a 8 site Hubbard model calculated using td-DMRG. A potential energy $U = 0.1t$ and a broadening $\eta = 0.1$ a.u have been used. Spectral functions have been calculated at the first site of the chain.

# 2 Dependence of the LDOS accuracy from $\tau$

A careful optimization of the $\tau$ value is fundamental to get accurate simulations and, at the same time, to avoid wasting time in excessively long propagations. In figure 2 the spectral function for the 8 site Hubbard model has been calculated using different values of the time-step. All the propagations have been carried out, in this case, for the same total period of time ($T = 500$ a.u.) using the RK4 scheme. As expected, reducing the size of the time-step (in particular in the 0.3-0.05 a.u. range) we are able to improve the quality of the spectrum, providing better and better approximations of the exponential propagator. Looking at these

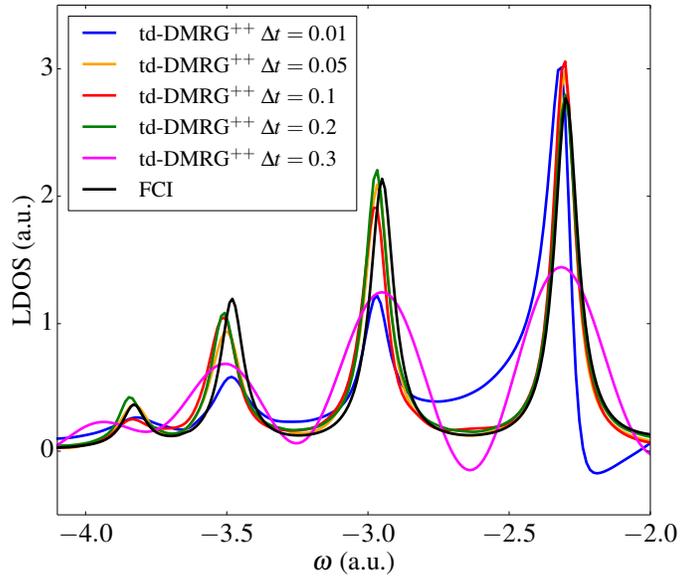

Figure 2: Dependence of the LDOS on the time-step ($\tau$) size. Calculations have been performed on an 8 site Hubbard model using the td-DMRG$^{++}$ approach. A potential energy $U = 0.1t$ and a bond dimension M=30 have been used. Spectral functions have been calculated at the first site of the chain.

results we can estimate a $\tau \approx 0.1 - 0.2$ a.u. as the best compromise between accuracy and computational cost. When smaller (e.g. 0.01 a.u.) time-step values are used, the results deteriorate (see blue line in figure 2). This behaviour was already observed by Feiguin and White in Ref. 1 and is attributed to an increase of the truncation error when large numbers of propagation steps are performed.

# 3 Additional DOS of Hydrogen Chains

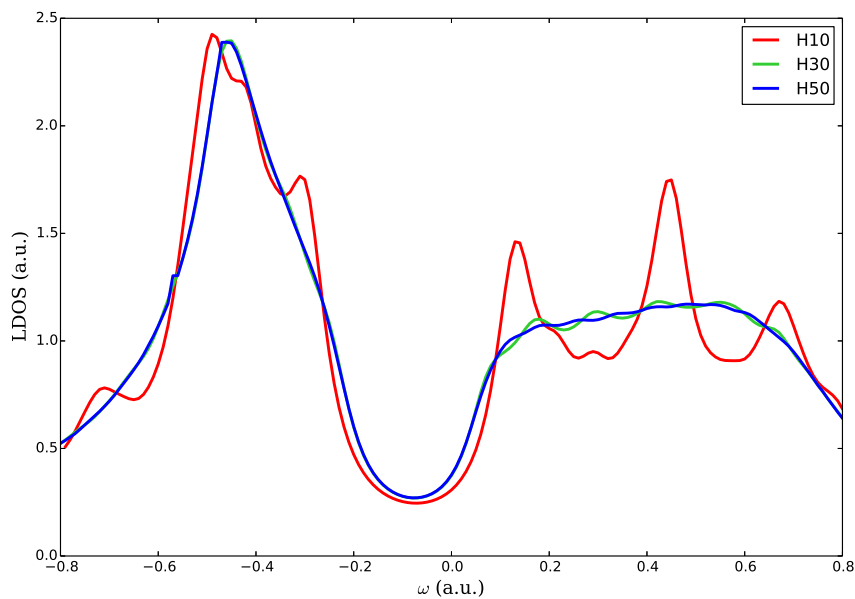

Figure 3: DMRG Spectral Functions of three equally spaced hydrogen chains ($H_{10}$ - red, $H_{20}$ - green, $H_{50}$ - blue) at r = 2.4 Bohr bond distance. All the LDOSs have been calculated on the central site of the chain. A broadening ($\eta$) equal to 0.05 a.u. has been used.

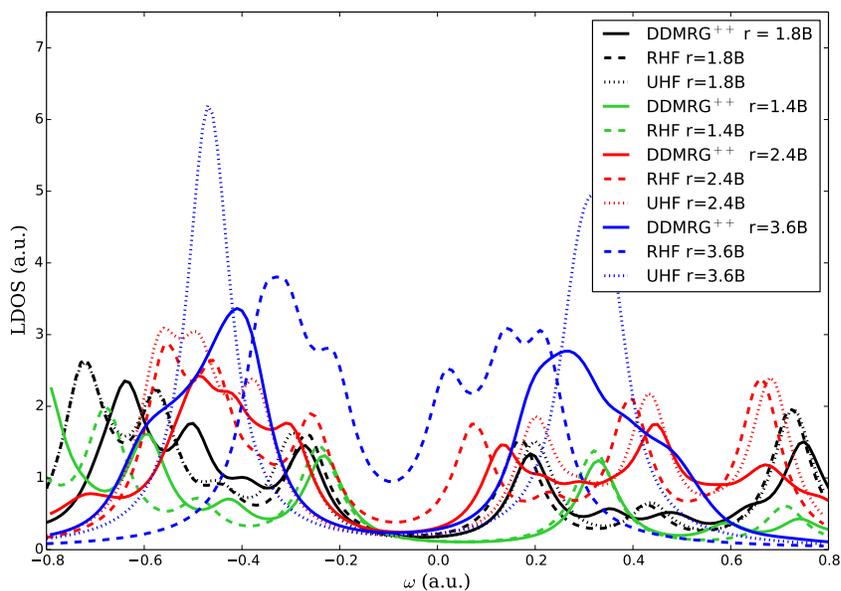

Figure 4: DMRG and HF Spectral Functions of a 10 atom hydrogen chain calculated at different inter-atomic distances. All the LDOSs have been calculated on the central site of the chain. A broadening ($\eta$) equal to 0.05 a.u. has been used.

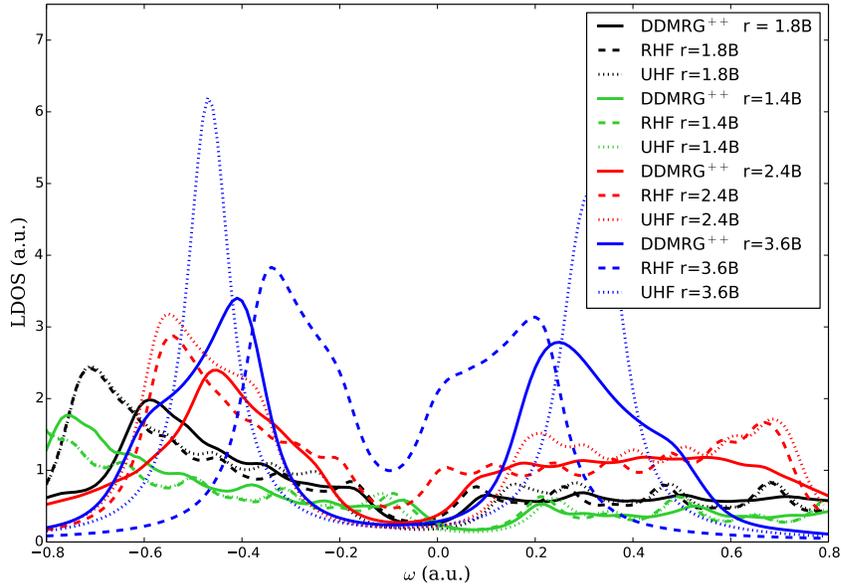

Figure 5: DMRG and HF Spectral Functions of a 30 atom hydrogen chain calculated at different inter-atomic distances. All the LDOSs have been calculated on the central site of the chain. A broadening ($\eta$) equal to 0.05 a.u. has been used.

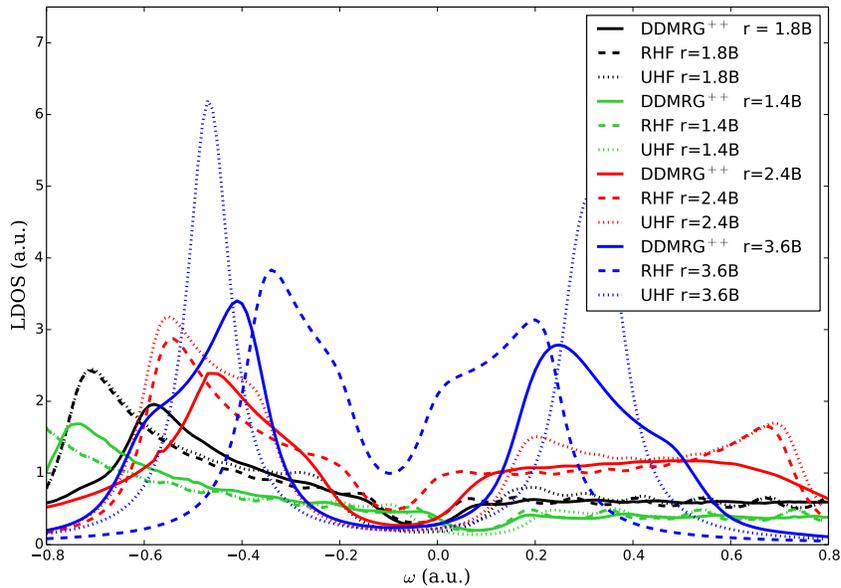

Figure 6: DMRG and HF Spectral Functions of a 50 atom hydrogen chain calculated at different inter-atomic distances. All the LDOSs have been calculated on the central site of the chain. A broadening ($\eta$) equal to 0.05 a.u. has been used.



\end{document}